
\magnification=1200
\hsize=6.4 truein \vsize=8.9 truein

\tolerance=10000
\def\folio{\ifnum\pageno=1\nopagenumbers\else\number\pageno\fi}
%
        
        \skip\footins=30pt
        \def\footnoterule{\kern-3pt
        \hrule width \the\hsize \kern 2.6pt}
\baselineskip 12pt minus 1pt
\parskip=\medskipamount
\def\nulltest{}    
\def\headline{}    
\def\headlineb{}   
\def\makeheadline{\ifnum\pageno=1 \else
  \baselineskip 12pt \ifx \headline \nulltest \else
     \line{\headline} \advance \vsize by -\baselineskip \advance
     \vsize by -14pt \fi
   \ifx \headlineb \nulltest \else
     \line{\headlineb} \advance \vsize by -\baselineskip \fi
   \ifx \headline \nulltest \else \vskip 14pt \fi
  \fi}   

\def\refpar{\parshape=2 0truein 6.5truein 0.3truein 6.2truein}
\def\ref#1;#2;#3;#4{{\par\refpar #1, {\it #2}, {\bf #3}, #4}}
\def\prep#1;#2{{\par\refpar #1, #2}}
\def\book#1;#2;#3{{\par\refpar #1, {\it #2} #3}}

\def\ltsima{$\; \buildrel < \over \sim \;$}
\def\lsim{\lower.5ex\hbox{\ltsima}}
\def\gtsima{$\; \buildrel > \over \sim \;$}
\def\gsim{\lower.5ex\hbox{\gtsima}}

\def\kms{\,{\rm km\,s^{-1}}}

\def\pp{\dimen0=\hsize
        \advance\dimen0 by -1truecm
        \par\parshape 2 0truecm \dimen0 1truecm \dimen0 \noindent}

\def\ltsima{$\; \buildrel < \over \sim \;$}
\def\lsim{\lower.5ex\hbox{\ltsima}}
\def\gtsima{$\; \buildrel > \over \sim \;$}
\def\gsim{\lower.5ex\hbox{\gtsima}}
\vbox to 0.15in{ }
\vskip .2truein
\centerline{\bf COLD DARK MATTER II:}
\centerline{\bf SPATIAL AND VELOCITY STATISTICS}
\vskip .15truein
\centerline{\bf James M. Gelb$^\dagger$ and Edmund Bertschinger}
\vskip .15truein
\centerline{Department of Physics,}
\centerline{Massachusetts Institute of Technology,}
\centerline{Cambridge, MA 02139}
\vskip .15truein
\centerline{$^\dagger$Present address:}
\centerline{NASA/Fermilab Astrophysics Center,}
\centerline{Fermi National Accelerator Laboratory,}
\centerline{P.O. Box 500, Batavia, IL 60510}
\vskip .15truein
\centerline{\bf ABSTRACT}
\vskip .15truein

We examine high-resolution gravitational N-body simulations of
the $\Omega=1$ cold dark matter (CDM) model in order to determine
whether there is any normalization of the initial density fluctuation
spectrum that yields acceptable results for galaxy clustering and velocities.
Dense dark matter halos in the evolved mass distribution are identified
with luminous galaxies; the most massive halos are also considered as
sites for galaxy groups, with a range of possibilities explored for the
group mass to light ratios.  We verify the earlier conclusions of White
et al. (1987) for the low amplitude (high bias) CDM model --- the
galaxy correlation function is marginally acceptable
but that there are too many galaxies.  We also show that the peak biasing
method does not accurately reproduce the results obtained using dense halos
identified in the simulations themselves.  The COBE anisotropy implies a
higher normalization, resulting in problems with excessive pairwise galaxy
velocity dispersion unless a strong velocity bias is present.  Although we
confirm the strong velocity bias of halos reported by Couchman \& Carlberg
(1992), we show that the galaxy motions are still too large on small scales.
We find no amplitude for which the CDM model can reconcile simultaneously
the galaxy correlation function, the low pairwise velocity dispersion, and
the richness distribution of groups and clusters.  With the normalization
implied by COBE, the CDM spectrum has too much power on small scales if
$\Omega=1$.
\vskip .3truein
\noindent{\it Subject headings:}
cosmology: theory --- dark matter --- galaxies: clustering ---
galaxies: formation

\vskip .2truein
\vfill
\eject
\vskip .3truein
\centerline {\bf 1.~INTRODUCTION}

The cold dark matter (CDM) model of galaxy formation has had a
checkered history.  First proposed by Peebles in 1982, the model has
the virtue of being relatively well-defined and testable.  Assuming
$H_0=50\,{\rm km\,s}^{-1}\,{\rm Mpc}^{-1}$ and $\Omega=1$, the only
fundamental free parameter is the overall amplitude of density
fluctuations.  We characterize this by the conventional quantity
$\sigma_8$, defined to be the rms density fluctuation, using the
linear power spectrum, in a sphere of radius $800\,{\rm km\,s}^{-1}$.
(See, e.g., Bertschinger 1992 for discussion of this and alternative
conventions for the normalization.)
Once $\sigma_8$ is specified the CDM model has, in principle, strong
predictive power, although many of the predictions require N-body and
dissipative numerical computations.  The complexity of the nonlinear
evolution and dissipation has led over the last decade to a lively debate
concerning the viability of the CDM model.

In 1985, Davis et al. showed that the CDM model cannot simultaneously
fit galaxy clustering (i.e., two-point correlation function) and small-scale
velocities (i.e., pairwise velocity dispersion) for any $\sigma_8$ if
dark matter traces galaxies.  For $\sigma_8=1$, motivated by the observation
that $\sigma_8\approx1$ for galaxies (Davis \& Peebles 1983), the relative
velocities of galaxies are predicted to be much larger than observed.  The
solution proposed by Davis et al. was to decrease the amplitude of density
fluctuations by a factor 2.5 to $\sigma_8=0.4$, thereby decreasing the
pairwise velocity dispersion of galaxies to roughly match observations.
In the process, the clustering of galaxies was also diminished.  The
two-point correlation function was boosted up to the observed range by
assuming that galaxies form only in the initially highest-density regions,
according to the ``peak biasing'' scheme proposed by Kaiser (1984).  Roughly
speaking, galaxy density fluctuations were assumed to be 2.5 times the dark
matter fluctuations (although peak biasing does not give an exactly linear
relation between galaxies and mass fluctuations).  The stronger correlations
introduced with biasing compensated for the smaller correlations (and
velocities) resulting from lowering the amplitude to $\sigma_8=0.4$.
The paper of Davis et al. (1985, hereafter DEFW)
marked the birth of the biased CDM model.

The amplitude of the CDM density fluctuations affects large-scale structure
as well as small-scale (20 Mpc and less) clustering and velocities.  Many
authors have pointed out that the low amplitude of biased CDM is in apparent
conflict with large-scale structure.  The freedom to vary the biasing
(ratio of galaxies to mass) makes it somewhat difficult to pin down these
problems.  However, large-scale peculiar velocities are particularly useful
for testing the normalization because galaxies should trace the same
large-scale flows as does dark matter --- all bodies accelerate the same
way in a gravitational field.  Motivated by this fact, Bertschinger \&
Juszkiewicz (1988) tested CDM predictions against the ``great attractor''
fits of Faber \& Burstein (1988) and concluded that CDM with $\sigma_8
\le 2/3$ was inconsistent with the data.  This conclusion was strengthened
by analysis of several large-scale galaxy surveys: the angular correlation
function of Maddox et al. (1990) from the APM survey; the moments of galaxy
counts in cells of Saunders et al. (1991) from the IRAS/QDOT survey; and the
galaxy density power spectrum from the CfA2 redshift survey by Vogeley et
al. (1992).

The large-scale structure measurements indicate a need for more power,
hence a larger $\sigma_8$ if the CDM power spectrum is retained.  However,
high-amplitude CDM faces the problem of large velocity dispersion noted
originally by Davis et al. (1985).  A possible solution was suggested by
Carlberg \& Couchman (1989) and Carlberg, Couchman, and Thomas (1990):
velocity bias.  They noted that the pairwise velocity dispersion of dark
matter halos in high-resolution N-body simulations is substantially less
than that of the mass.  This effect was not discovered by Davis et al.
because their simulations did not have the resolution needed to find
galaxy halos composed of many particles (although it could have been
found by White et al. 1987).  Couchman \& Carlberg (1992)
pointed out that with an amplitude corresponding to $\sigma_8=1$, CDM
would do well on large scales, while clustering and velocity biasing
might solve the problems on small scales.

However, it is not enough for CDM to predict the correct two-point
correlation function and pairwise velocity dispersion of galaxies.
It must also predict the correct abundance of galaxies and of galaxy
groups as a function of richness.  Testing these requires high-resolution
numerical simulations.  An important first step was taken in 1987 by
White et al. (hereafter WDEF).  They performed a  P$^3$M simulation,
evolved to $\sigma_8=0.4$, in a 50 Mpc box with enough particles ($64^3$)
to study resolved dark matter halos.  They found that the evolved halos
are, indeed, more strongly correlated than the mass, with the correlation
function in reasonable agreement with observations for halos with circular
rotation speeds exceeding $250\kms$.  However, they found the numbers of
halos in their simulations with $V_{\rm circ}\ge 100\kms$, after breaking
up the overly-merged massive halos, to be greater than the observed numbers
by a factor of 2 or more.

The uncertainty over the correct amplitude for normalizing the CDM (or any
other) power spectrum has largely ended with the measurement of the cosmic
microwave background anisotropy by Smoot et al. (1992).  Their measurements
imply $\sigma_8 \approx 1.1$ (Wright et al. 1992; Efstathiou, Bond, \&
White 1992; Adams et al. 1993) if the spectrum is scale-invariant as we
suppose.  Consequently, all standard CDM models with $\sigma_8=0.4$ are
obsolete.  On the other hand, CDM with $\sigma_8=1$ might be an attractive
model if the small-scale velocity bias found by Carlberg et al. is
sufficiently strong and if the galaxy numbers and group multiplicities
can be made reasonably to match the observations.

In previous work (Bertschinger \& Gelb 1991)
we have also found a strong
velocity bias for halos in the CDM model.  However, our interpretation
differs somewhat from that of Carlberg (1991); the reduction appears
to be a statistical effect, arising because the pairwise velocity statistic
weights galaxies by pairs and quadratically by velocities.  Therefore, when
it is applied to all mass particles, the massive halos in galaxy clusters,
with velocities comparable to the cluster dispersion, contribute strongly
to the overall pairwise velocity dispersion.  When the simulated halos are
used, on the other hand, strong merging in the clusters eliminates most of
the halos and leaves a single massive object in the center, which has
little weight in the pairwise sum.  The small number of halos in clusters
reduces the pairwise velocity dispersion of halos, but it also leads to
clusters with far too few objects to be identified as galaxies.

In a preceding paper (Gelb \& Bertschinger 1993, hereafter Paper I),
we explored in detail the distribution of simulated halos by circular
velocities, and we concluded that no value of $\sigma_8$ could fully
satisfy the constraints given by the number density of galaxies.  For
any value of $\sigma_8$ one had too many halos, assuming that circular
velocity can be related to luminosity by Tully-Fisher and Faber-Jackson
relations.  However, these results are limited by the fact that the
N-body simulations we used lack gas, so it is worthwhile making other
tests of the model, with generous allowance for uncertainties in how
galaxies are related to dark matter halos.

Dissipationless simulations like ours have the disadvantage of being
unable to correctly model the process of galaxy formation inside dark
matter halos.  By necessity, we assume that galaxies form only inside dark
matter halos.  This assumption appears to be confirmed by gas dynamical
simulations (e.g., Katz, Hernquist, and Weinberg 1992; Cen \& Ostriker
1992a,b; Evrard, Summers, \& Davis 1994),
which also support our conclusions concerning the degree
of velocity biasing.  Why then should we continue with dissipationless
simulations?  The reason is dynamic range.  With simulations including
gravity only, we are able to resolve galaxy halos in volumes up to 100 Mpc
on a side, large enough to capture the long wavelength density fluctuations
important for high-amplitude CDM (cf. Paper I).  Gas dynamical simulations
with equal numbers of particles are still prohibitively expensive.  The
volumes studied to date with high resolution gas dynamical simulations
have been too small to include all important dynamical effects.

In this paper we study spatial clustering and velocity statistics of dark
matter halos in the CDM model using high-resolution N-body simulations.
The principle goal is to answer the question:
is there a normalization $\sigma_8$ such that the two-point
spatial correlation function of the resolved halos and the pairwise
velocity dispersion of the resolved halos matches the observations?
This question is addressed by analyzing the particle-mesh and
particle-particle/particle-mesh (P$^3$M) N-body simulations discussed
in paper I.
For economy of notation (see Paper I) we refer to the simulations
as CDMn($N^3$,$L$,$R_{1/2}$).
The numbers in parentheses indicate the following simulation parameters:
1) $N^3$ particles, 2) a comoving box of length $L$ Mpc on a side,
and 3) a comoving force softening length of $R_{1/2}$ kpc.
The force softening is characterized by $r=R_{1/2}$:
where $r^2F_r/(Gm^2)=1/2$, i.e. half its Newtonian value.

Our studies focus on the simulation CDM16($144^3$,100,85).
The comoving Plummer softening length is $\epsilon=65$ kpc
and the particle mass is $m_{\rm part}=2.3\times 10^{10}{\rm M}_\odot$.
Of all our simulations (see Gelb 1992, Appendix II), this simulation
has the best compromise of mass and force
resolution in a 100 Mpc box.  A relatively large box
is required to adequately represent waves in the initial conditions,
particularly for evolution up to $\sigma_8=1$
(see Paper I and $\S\S$ 2.1 and 2.2 below).
We will examine positions and peculiar velocities at several amplitudes:
$\sigma_8=0.4$, 0.5, 0.7, and 1.0.  Each epoch studied is considered a
candidate for the present day; i.e. we test whether spatial and velocity
statistics match the observations.  The box length is assumed to have a
physical length of 100 Mpc for CDM16 with a Hubble constant
$H_0=50\kms~{\rm Mpc}^{-1}$ (and $\Omega=1$) at each candidate epoch.

We will begin by exploring some background material: box size, the
standard CDM model, massive halos, and velocities in $\S$ 2.
We then briefly discuss some limitations of the method of peak
particles and argue the necessity of using resolved halos
to study the CDM model in $\S$ 3.
In $\S$ 4 we study the two-point
correlation function of simulated galaxies, while in $\S$ 5 we study
the abundance and richness distribution of galaxy groups.  In both
cases the results depend on how over-merged halos are broken apart,
a point that we investigate in some detail.  In $\S$ 6 we investigate
the small-scale velocity statistics of the galaxies.  In $\S$ 7 we summarize
the implications for the $\Omega=1$ CDM model.
Further numerical details can be found in
Bertschinger \& Gelb (1991), Gelb (1992), and Paper I.
\vfill
\eject
\vskip .3truein
\centerline {\bf 2.~BACKGROUND AND THE STANDARD MODEL}

In this section we explore the issue of box size;
we test whether we can reproduce the results of White et al. (1987);
and we consider complications arising from massive
halos and from definitions of halo velocities.
We will confirm that the ``standard'' biased CDM model
($\Omega=1$, $\sigma_8=0.4$)
produces far too many halos compared with
the observations.

\vskip .2truein
\centerline {\it 2.1.~Box Size: Mass Correlation Length}

We are interested in studying models evolved to $\sigma_8=1$.
More highly evolved models require larger boxes since successively larger
waves begin to go nonlinear.  In this subsection
we examine linear theory predictions regarding the importance
of long waves in the initial conditions.  We then present
nonlinear results using evolved
$N$-body simulations to study the dependence on box size
of the two-point correlation function and the
pairwise velocity dispersions of the mass.

In Figure~1a we show the linear, rms
mass fluctuation in a sphere of radius $8h^{-1}=16$ Mpc.
We normalize the Holtzman (1989) (5\% baryons) CDM
power spectrum so that $\sigma_8=1$.
We then compute $\sigma_8(\lambda_{\rm max})$ by including only waves
with wavelength less than $\lambda_{\rm max}$ in the numerical integration
of $\sigma_8$.
The value of the maximum wavelength represented in a simulation computed
in a box of length $L$ on a side is $\lambda_{\rm max}=L$.
As $\lambda_{\rm max}\rightarrow \infty$,
$\sigma_8(\lambda_{\rm max})\rightarrow 1$ by definition.
We show vertical bars at $\lambda_{\rm max}=$ 51.2 Mpc and 100 Mpc,
the sizes of several of our simulation cubes.

For $\lambda_{\rm max}=51.2$ Mpc we find $\sigma_8(\lambda_{\rm max})\approx
0.6$.
For $\lambda_{\rm max}=100$ Mpc we find $\sigma_8(\lambda_{\rm
max})\approx 0.9$.  For $\lambda_{\rm max}=150$ Mpc we find
$\sigma_8(\lambda_{\rm
max})\approx 0.98$ and it quickly approaches unity thereafter.
We conclude that long waves (with $\lambda > 50$ Mpc) make
significant contributions to $\sigma_8$ and, by extension, to
the correlation length $r_0$ (where the two-point correlation function
$\xi=1$).  Therefore we expect that the nonlinear evolution of our
$N$-body simulations up to the amplitude $\sigma_8=1$
will result in a serious underestimate of $r_0$ in 51.2 Mpc boxes but
probably not in 100 Mpc boxes.

To test the above prediction, in Figure~2 we show $\xi(r)$
for various simulations for the mass (particles).
The important parameter for our discussion is the box size.
The plots show that the correlation length $r_0$ grows roughly linearly
with $\sigma_8$ from 5 Mpc at $\sigma_8=0.5$ to 10 Mpc at $\sigma_8=1$ for the
simulations in larger boxes.  We must match the observations with
the simulated halos, not the mass, unless the halos
trace the mass.   However, we expect the same waves that
affect the mass will also affect halo clustering.
We notice in Figure~2 that the 51.2 Mpc box simulations
underestimate the correlation length (by about 30\% for $\sigma_8=1$)
when compared with the simulations in $\ge$ 100 Mpc boxes.
The underestimate is greater for increasing $\sigma_8$ (being about
20\% for $\sigma_8=0.5$), as expected.  Boxes smaller than 50 Mpc
on a side are too small to get the correlation length correct to
better than 10\% even for an amplitude as small as $\sigma_8=0.4$.
However, we find that a 100 Mpc box is sufficiently large, as the
correlation length (10 Mpc) agrees with the results from the 400 Mpc box
simulation, despite the poor resolution (2 Mpc) of the latter.

{}From Figure~2 we can also examine the
the simulation-to-simulation variation in the value of $r_0$.
We see basic agreement in $r_0$ among the larger box simulations.
However, we do see a smaller value in $r_0$ at $\sigma_8=0.7$
for CDM16 (100 Mpc box) indicating that there are simulation-to-simulation
fluctuations.  The fluctuations for the five $R_{1/2}=280$ kpc
PM simulations in 51.2 Mpc boxes are shown
with $1\sigma$ error bars.  The largest fluctuations are found
on the largest scales.

We conclude that
the 51.2 Mpc boxes are too small for accurate predictions of
the two-point correlation function particularly for
larger values of $\sigma_8$.
It is unfortunate
that CDM16 (100 Mpc box) has a slight sag on scales near 10 Mpc
owing to a statistical fluctuation.
This will limit some of the conclusions we can draw
from this single simulation.  Nevertheless, the large numbers of halos
in this large box provide a fair test of the CDM
model.  More realizations would be helpful, but
our other $\sim 100$ Mpc box simulations have
poor force and mass resolution.

\vskip .2truein
\centerline {\it 2.2.~Box Size: Mass Pairwise Velocity Dispersion}

In this subsection we examine the effect of the box size on the
pairwise velocity dispersion $\sigma_{\rm p}(r)$
as a function of galaxy separation:

\vbox{
$$\sigma^2_{\rm p}(r)\equiv
{1\over3}\,\langle (\vec v_2 - \vec v_1)^2 \rangle\ ,
\eqno(2.1)$$}
{\parindent 0pt
where the average is taken over pairs of particles (for the mass) or
halos separated by distance $r=\vert\vec r_1-\vec r_2\,\vert$,
with peculiar velocities $\vec v_1$ and $\vec v_2$.
Nonlinear studies of velocity dispersions are the subject
of Gelb, Gradwohl, \& Frieman (1993).}

For an initial estimate of effects of finite box size we consider the
linear theory prediction for the three-dimensional pairwise dispersion
$\sigma_v(r)=3^{1/2} \sigma_{\rm p}(r)$.
We evaluate this quantity using the $\sigma_8=1$ linear normalization,
with a wavelength cutoff $\lambda_{\rm max}$ applied to the numerical
integration (as we did in $\S$ 2.1) to mimic the effect of a finite box
size.  The results are shown in Figure~1b.
We see there is a significant difference between a
25 Mpc box and a 50 Mpc box (particularly for larger
values of $r$) and a smaller difference
between a 50 Mpc box and a 100 Mpc box.  The results
converge on the scales of interest in this paper, i.e.
$r\sim 1-10$ Mpc,
for $\lambda_{\rm max}\gsim 100$ Mpc.

In Figure~3 we show
$\sigma_{\rm p}(r)$ for the mass for the same simulations
studied in $\S$ 2.1.
We notice that the velocities
are higher for the simulations in boxes of size $\ge$ 100 Mpc
than for the simulations in 51.2 Mpc boxes --- long waves in the initial
conditions affect nonlinear pairwise velocities on smaller scales (see also
Gelb, Gradwohl, \& Frieman 1993).
We see that 100 Mpc boxes are sufficient because
the results for the 400 Mpc box simulation are comparable
to the other 100 Mpc box simulations.  The 400 Mpc box simulation
has extremely soft forces
($R_{1/2}$=2 Mpc) so the velocities are actually
lower than in the 100 Mpc simulations.
(The parameters in the 400 Mpc simulation are close
to the values used by Park 1990.)  We use our results from
linear theory to strengthen our argument that
a 100 Mpc box is sufficient for studying
the velocities.  This is encouraging because we demonstrated
in the previous section that a 100 Mpc box is
sufficient for studying spatial correlations.

\vskip .2truein
\centerline {\it 2.3.~The Standard Biased CDM Model: $\sigma_8=0.4$}

In the above sections we found that a box of size 51.2 Mpc is too
small to include all the long-wavelength contributions important for
clustering and pairwise velocity dispersions on small scales.  However,
much of the past work has been done using simulations of this size or
smaller.  To show that our simulations are consistent with previous
work, we perform two simulations with parameters similar to
those used by WDEF.  These P$^3$M simulations, CDM12 and CDM13, use
$64^3$ particles in 51.2 Mpc boxes with Plummer softening $\epsilon=40$
kpc comoving.  The particle mass is $m_{\rm part}=3.5\times 10^{10}\,
{\rm M}_\odot$.  For comparison, WDEF computed three $64^3$ particle
P$^3$M simulations in 50 Mpc boxes with force resolution $\sim50$
kpc comoving (they used a linear sphere density profile).
We study the models at $\sigma_8=0.4$, the
normalization advocated by DEFW.  To identify dark matter halos we
use two different prescriptions: DENMAX (our density maxima finder,
see Paper I) and FOF (friends-of-friends); WDEF used the latter.
For friends-of-friends we report the linking length, $l$,
in units of the mean interparticle spacing.
We study the numbers of simulated halos and
we break up the massive halos into clusters to
study the effect on the spatial clustering of massive
halos with circular velocities $V_{\rm circ}\ge 250\kms$.

In Table~1 we list the numbers of halos with
$V_{\rm circ}\ge 100,~200,~$ and 250 $\kms$.  The
results are shown for averages from CDM12($64^3$,51.2,52) and
CDM13($64^3$,51.2,52) (two different sets
of initial random numbers) and for averages from
WDEF. (We divide their numbers by three since they report totals
for three simulations.  We also scale their numbers in a 50 Mpc
box to a 51.2 Mpc box.)
In computing observational estimates we characterize all our halos
by their circular velocities, and we relate observed estimates of
one-dimensional velocity dispersions, $\sigma_1$,
to circular velocities by
$$\sigma_1=F{V_{\rm circ}\over{\sqrt{3}}}~.\eqno(2.2)$$

We report our observational estimates
for $F=1$ and $F=1.1$ (see Paper I) and those of
WDEF for $F=1$ (in our notation).  WDEF
used $F=1$ but in a subsequent paper
at $\sigma_8=0.4$ they used $F=1.1$ (see Frenk et al. 1988).
We demonstrated in Paper I that $F=1.1$ works better
for $\sigma_8=0.5$ compared with larger values of $\sigma_8$.
In any event, $F=1.1$ lowers the observational estimates making
the disparity with the observations worse.
The observational estimates of WDEF are higher than ours.
This stems from the fact that they used a different
Faber-Jackson relationship for ellipticals.
We note that the Faber et al. (1989) survey of ellipticals has even
fewer bright ellipticals (factor $\sim 1/2$ for $\sigma_1\ge 350\kms$)
than we get with our Faber-Jackson relation,
$M_{B_T}=-6.6364\log_{10}(\sigma_1)-5.884$.
The differences are not critical
since we will find, using either our estimates or those of WDEF,
that there are too many simulated halos compared
with the observations.  WDEF study the clustering of
halos with $V_{\rm circ}\ge 250\kms$ but they only
report numbers for $V_{\rm circ}\ge 100$ and 200 $\kms$,
so the entry for $250\kms$ is unfilled in Table~1.

The triplet of numbers in Table~1 are for
$V_{\rm circ}\ge 100,~200,~250\kms$.  Results are shown for
halos defined using
FOF and DENMAX, and for DENMAX after breaking up the massive halos
into groups and clusters of halos (see $\S$ 4.2 and 4.3).
The WDEF results are shown before and after their special treatment
of merging.  Observational estimates are given in the first column.

We identify halos in our simulations with FOF linking lengths
$l=0.1$ and 0.2, and DENMAX with a $512^3$ density grid.
Circular velocities are defined at a comoving radius of 150 kpc.
We find similar results using different DENMAX grids
and different radius cuts (see Paper I).  The variation
in the numbers from CDM12 versus CDM13 are less than 10\%.
We see from Table 1 that the results for the DENMAX analyses
are closer to the FOF ($l$=0.1) case than the FOF ($l$=0.2) case
for the larger circular velocity cut-offs.
WDEF reported their results before and after their prescription
for restoring merged halos into clusters.  They used FOF with
an unspecified, small linking parameter and
they defined their circular velocities using a mass within
a sphere of mean density 1000 times the present critical density.
Despite these differences, we find reasonable agreement for
the numbers of halos with $V_{\rm circ}$ exceeding $200\kms$.
However, their break-up procedure results in more than twice as
many halos as we find for $V_{\rm circ}\ge100\kms$.  In any case,
the simulations predict more than twice as many halos as there
ought to be.

In Figure~4
we show averages of the two-point correlation functions from
the two P$^3$M simulations for the
halos with $V_{\rm circ}\ge 100$, 200, and 250 $\kms$ at
$\sigma_8=0.4$.  The mass is shown as a solid curve
(also with $1\sigma$ error bars in the bottom panel).
Our mass correlation length agrees with WDEF: $r_0\approx 4$ Mpc.
We show the observed two-point correlation function ($r_0=10$ Mpc;
logarithmic slope $-1.8$) as a straight solid line.
We see
a slight enhancement of the correlation length of the halos (dotted and
dashed curves)
compared with the correlation length of
the mass.  WDEF reported a similar enhancement.
The two-point correlation function also has the wrong shape;
in agreement with WDEF it is too steep on small scales and
has a sag at $r\sim 1.5$ Mpc.

The enhancement in $\xi$
for the bright halos is not large enough to reconcile
$\sigma_8=0.4$ with the observations.  WDEF argued,
and we present more arguments later, that the
massive halos might represent clusters of galaxies that
have merged.  WDEF, in a complicated manner, found
every halo that ever formed in the evolution of their
models and then used a prescription for merging.
(Rather than mimic their procedure in detail, we simply
add halos to each massive system in proportion to its bound mass.)
When WDEF applied their algorithm to their models,
they found significantly more sites for
galaxy formation compared with not breaking up the massive halos.
(We list their numbers as {\it before} and {\it after}
in the break-up column of Table~1.)
They found $\xi$ is significantly enhanced
after break-up, and halos with $V_{\rm circ}\ge 250\kms$ then matched
the observed $\xi$ fairly well.  If one adds halos to massive systems,
one gives extra weight to these systems which are more correlated
than smaller systems, thereby increasing the correlation function
(Kaiser 1984).

We show our results with the break-up of massive halos as points
in the bottom panel of Figure~4.  We show
$1\sigma$ error bars from the two simulations.  This procedure
introduces new small-scale pairs enhancing $\xi$ on small scales.
We find for one of our simulations
that $\xi$ comes close to the observed line and that for the other
simulation the enhancement at large $r$ is small.  We find
that $r_0$ for the mass has a substantial simulation-to-simulation
fluctuation at $r\sim 10$ Mpc which is not surprising in a small
box (see Figure~2).  The larger correlation length of the
halos corresponds to the simulation with the larger correlation
length in the mass.  Giving extra weight to the massive halos
enhances $\xi$ provided that there are significant contributions to $\xi$
from long wavelengths.

Our crude break-up scheme
is less ambitious than that of WDEF and it produces fewer halos.
We include it here only to illustrate the points made by
WDEF and to emphasize the problem associated with producing too many
halos.  In agreement with WDEF, we find
that it is essential to break up the massive halos
in order to approximately reconcile the two-point correlation function
of resolved halos at $\sigma_8=0.4$
with the observed two-point correlation function.
However, if we break up the massive halos
enough to enhance $r_0$ to match the observations, then
the resulting numbers of halos far exceed the observational
estimates.  Moreover
$\xi(r)$ still has the wrong shape.  These facts must be considered
to be serious shortcomings of the model.  The discrepancies must
be considered tentative, however, because we know that a 51.2 Mpc
box is too small.  We use a 100 Mpc box in subsequent sections.

\vskip .2truein
\centerline {\it 2.4.~Halo Velocities}

In this subsection we present a few comments concerning the
velocities of resolved halos.  This subject is important for
assessing velocity bias:
the pairwise velocity dispersion of the halos can much be less
than that of the mass (see Carlberg, Couchman, \& Thomas 1990).
In Figure~5 we show $\sigma_{\rm p}(r)$
at $\sigma_8=0.7$ for CDM12($64^3$,51.2,52)
for the mass and for the halos (found with a 512$^3$ DENMAX grid).
We use the center-of-momentum to define the velocities of the halos.
We see that $\sigma_{\rm p}$ is significantly smaller for
the halos than for the mass.
One possible source of this ``velocity bias'' is dynamical friction
of the halos
with the surrounding medium (Carlberg, Couchman, \& Thomas 1990),
where the internal motions of particles in the halo exchange
energy and momentum with surrounding particles or other halos.

Much of the velocity bias, however, might actually be a statistical effect.
To see this, in Figure~5 we also show $\sigma_{\rm p}$ for the mass with the
particles from the two largest halos removed (dashed-dotted curve).
These large halos have  circular velocities
(defined with $R=150$ kpc comoving) of 981 $\kms$ and 904 $\kms$.
We remove all of the DENMAX particles---even the unbound ones---
which involve 7719 particles and 4603 particles respectively.
Removal of these particles is unrelated to dynamical friction.
We see that the mass pairwise velocity dispersion is reduced significantly
by this removal, and therefore the amount of dynamical friction required to
explain the overall velocity bias is less than one might expect.
The reason why removing the massive halos has such a large effect
on $\sigma_{\rm p}$ is simple (Bertschinger \& Gelb 1991).
The calculation of $\sigma_{\rm p}$ for the particles weights each pair,
giving quadratically greater weight to pairs of the most massive halos.
The pairwise velocity dispersions of these large objects are also much
higher than they are for smaller systems.  If we remove large halos we
remove a large number of pairs of high-velocity particles.

We now consider the important distinction between using
the velocity of the maximally bound particle from a halo
(used by WDEF) and the center-of-momentum velocity.
The short dashed curve in Figure~5 is
$\sigma_{\rm p}$ for the halos with $V_{\rm circ}\ge 192\kms$,
but we use the velocity of the maximally bound particle;
i.e. the one with the minimum potential computed by direct summation
of particles in the halo treated in isolation.
By using the center-of-momentum for the
velocity of the halo rather than the velocity of
the maximally bound particle, we get lower values
of $\sigma_{\rm p}$ because individual particles have
a significant velocity dispersion about the mean halo velocity; i.e.
we are not including the
internal halo velocity dispersion when we use the center-of-momentum.
It makes sense to define the velocity of a halo using the
center-of-momentum of the halo because observers define redshifts
using the average velocities of the stars in a galaxy.

\vskip .3truein
\centerline {\bf 3.~PEAK PARTICLES}

In this section we discuss an alternative definition of ``galaxies''
based on particles initially in density peaks.  This method is often
used with N-body simulations lacking sufficient mass or force resolution
to resolve evolved dark matter halos (see Kaiser 1984; DEFW; Bardeen
{et al.} 1986; Park 1990; Park 1991; Katz, Quinn, \& Gelb 1993).
Galaxies are identified as particles nearest initial (linear) density
maxima and their evolution is followed along with the other particles
representing intergalactic clouds of dark matter.  Only peaks with
density exceeding some threshold are accepted.  The peak threshold,
for a given gaussian smoothing radius used to smooth the initial
density field, is chosen to give the correct number of bright
halos in the simulation volume.

\vskip .2truein
\centerline {\it 3.1.~The Two-point Correlation Function: $\xi(r)$}

Following Frenk et al. (1988), we use gaussian smoothing radii $R_s$
of 550 kpc comoving and 880 kpc comoving, with corresponding density
fluctuation thresholds $\nu=2.6$ and 3.0 (for $R_s=550$ kpc comoving)
and $\nu=2.0$ and 2.5 (for $R_s=880$ kpc comoving) in units of $\sigma_\rho$---
the density dispersion computed from the smoothed, initial density field.
These values of $\nu$ give roughly 700 and 1600 galaxies, respectively,
in a $(100~{\rm Mpc})^3$ volume (our simulation CDM16).
Using the parameters in the Schechter luminosity function
(see Paper I where we used parameters from Efstathiou, Ellis, \&
Peterson 1988) these correspond to circular velocity cut-offs of approximately
$250\kms$ and $200\kms$ respectively.  For a given gaussian smoothing radius
smaller values of $\nu$ correspond the smaller circular velocity cutoffs.
We choose two values of $R_s$ to test the sensitivity of our results
to this parameter.

We next determine $\xi$ computed using only the peak particles in the
CDM16 simulation.  In other words, we compute the two-point correlation
function using the present positions of the particles
which are tagged as galaxies.  The results are shown in Figure~6.
The peak particles indicate that $\sigma_8=0.5$
is possibly suitable as the present epoch---the value of $r_0$
is roughly 10 Mpc and the logarithmic slope is very nearly
$-1.8$ from about 1 Mpc to 20 Mpc.  Note that this success is
exactly what led DEFW to champion biased CDM.  The correlation
lengths for $\sigma_8=0.7$ and $\sigma_8=1.0$ are also roughly 10 Mpc,
but the slope steepens at roughly $r\lsim$ 3 Mpc for $\sigma_8=0.7$
and at roughly $r\lsim$ 4 Mpc for $\sigma_8=1.0$.
Even for $\sigma_8=0.5$, the slope steepens for
roughly $r\lsim$ 1.5 Mpc.
(We also computed $\xi$ at $\sigma_8=0.4$ and
the results are nearly identical to $\sigma_8=0.5$ except
that the steepening of the slope occurs at $r\lsim 1.25$ Mpc
rather than $r\sim 1.5$ Mpc.)
The enhancement occurs because peak particles are more likely to
be found in massive halos where the chance of a peak
being above the threshold $\nu$ is higher.

To see how peaks are associated
with massive halos, we show in Figure~7 the bound particles from
a massive halo ($2.1\times 10^{14}~{\rm M}_\odot$)
at $\sigma_8=0.5$ from CDM16 (upper left panel).
We use a $512^3$ DENMAX grid which apparently has
not completely resolved all substructure.  We noted this
problem in Paper I where we concluded that increased force
resolution reveals substructure and increased DENMAX grids
are required to bring out this substructure.  However, in many
cases there is no obvious substructure in the images of the
massive halos.  In the upper right panel we
show peak particles that end up as bound members of the massive halo
shown in the upper left panel.
We see that there are many peak particles in this
massive halo.  (We discuss the other panels later.)

The large number of peak particles per massive halo is typical.
Conservation of numbers then implies that peak particles must
undersample less massive halos outside clusters.  To see this
quantitatively, we consider the resolved halos found by DENMAX
with $V_{\rm circ}\ge 250\kms$ defined at $R=200$ kpc comoving at
$\sigma_8=0.5$.  There are 737 halos.  Of these, we count the number
of halos that do not contain any peak particles as bound members.
For $R_s=550$ kpc comoving, using $\nu=3.0$ which yields
639 peaks, 425 resolved halos have no peak particles as bound
members.  (Using $\nu=2.9$ yields 826 peaks, and then 362
resolved halos have no peak particles as bound members.)
For $R_s=880$ kpc comoving, using $\nu=2.5$ which yields
740 peaks, 355 resolved halos have no peak particles as bound
members.  We conclude that about half of the massive halos
in the evolved, nonlinear density field contain no peak particles.
This is a major failing of the peak particles as galaxy tracers
(cf. Katz et al. 1993) and it calls into question
N-body simulations that rely on peak particles in lieu of dense
halos.

\vskip .2truein
\centerline {\it 3.2.~Discussion}

Should we conclude from Figure 6
that the correlation function slope in the CDM model is too steep on small
scales?  Park (1991) presented similar studies of $\xi(r)$ using peak
particles, but he did not show the steepening of $\xi$ within 1 Mpc.
His force resolution was of order $\sim 1$ Mpc (for a $256^3$ grid PM
simulation in a 153.6 Mpc box).  However, his models show, in agreement
with our results, the steepening of the slope at larger scales for
$\sigma_8=1.0$.  We will have to compute $\xi(r)$ using actual resolved
halos, with and without the break-up of merged massive halos, to decide
whether the steepening of $\xi(r)$ is real or an artifact of peak
particles.

In any case, we conclude that the method of peak particles can give
misleading results.  In $\S$ 2 we found that we could only get a significant
enhancement in $\xi$ at $\sigma_8=0.4$ if we broke up massive halos,
but this produced far too many halos.  We found that we can get the
required enhancement in $\xi$ at $\sigma_8=0.5$ (the results were similar
at $\sigma_8=0.4$) using the correct number of peak particles.  However,
half of the actual, massive, nonlinear halos did not contain any
peak particles.  Because there is nothing unusual about those halos
that do not contain peak particles (compared with halos of the same
circular velocity that do), we cannot argue that peak particles are to
be preferred over direct identification of dense resolved  halos.
Peak particles oversample the clusters and undersample the field.
These effects enhance the two-point correlation function with fewer halos.
For these reasons, we must study the CDM model using resolved dense
halos, rather than peak particles, to trace galaxies.

\vskip .3truein
\centerline {\bf 4.~TWO-POINT CORRELATIONS OF HALOS}

\vskip .2truein
\centerline {\it 4.1.~Introduction}

The preceding section has motivated a study of the two-point correlation
function using resolved halos, which we undertake in this section.
We use the high-resolution simulation CDM16($144^3$,100,85) in an attempt
to constrain $\sigma_8$ based on the slope and amplitude of $\xi(r)$.
We will find that the results depend on how merged halos are treated,
so we will devote some discussion to this issue.

We show $\xi$ computed from resolved halos at $\sigma_8=0.5$,
0.7 and 1.0 from CDM16 in Figure~8.
We use a $512^3$ DENMAX grid, with bound particles only (see Paper I),
for the remainder of this paper.  We see that the correlation length,
$r_0$, falls short of the observed value at low values of $\sigma_8$.
Also, there is increased merging at later epochs
(see Paper I); this explains why the halos are
antibiased, i.e. are less clustered than the mass, on small scales.
The antibiasing is stronger at later epochs and
for smaller halos.  This is because merging increases
with increasing $\sigma_8$ and the smallest halos merge
into larger systems.  Unfortunately for us, observers do not directly measure
the clustering of the mass; they measure the clustering
of the galaxies.  However, we see that unless galaxies are clustered
more strongly than the halos, we will not be able to match the observed
two-point correlation function.

Carlberg \& Couchman (1989) performed a simulation
with $1.2\times 10^{11}{\rm M}_\odot$ particles
in a 80 Mpc box.  (CDM16 has particles
with $2.3\times 10^{10}{\rm M}_\odot$ in a 100 Mpc box.)
At $\sigma_8=0.54$, using FOF to identify dark halos, they found,
as we do, that the dark halos are antibiased with respect to the
mass on small scales and that they trace
the mass on larger scales (see their figure 8b).
Couchman \& Carlberg (1992) studied more evolved models
and they also found the same level of antibiasing of the dark halos
with respect to the mass on small scales.

If we take these results for $\xi$ at face value, then
the $\Omega=1$ CDM model has serious shortcomings: the correlation length
is too small for $\sigma_8\lsim 0.7$; and the correlation
amplitude is too small and turns over on small scales,
particularly for $\sigma_8\gsim 0.7$.
Rather than abandon the model, however, we explore the possibility
that restoring merged halos might sufficiently increase $\xi$.
This step is reasonable, because the most massive halos clearly ought
to contain several galaxies.
However, we will find that we create as many problems
in the process as we solve!

We break up the massive halos using
two techniques.  The first method (fall-in method) involves finding the
maximally bound particle in each halo at an earlier epoch.  We identify
the maximally bound particles that have fallen into massive halos
representing clusters at the present epoch.  The second method
(mass-to-light method) involves assuming a mass-to-light ratio for
the massive halos representing clusters, and then, using the observed
Schechter luminosity function, assigning the appropriate number of members
to the clusters.  The fall-in method directly shows that smaller halos merge
into larger halos, as we showed in Paper I.  Perhaps dissipative effects
(Katz et al. 1992; Katz \& White 1993;
Evrard et al. 1994) or harder forces (Dubinski \& Carlberg 1991)
might help these systems survive the merging process.  However, because
clusters exist but are over-merged in our simulations, it is reasonable
to unmerge the most massive halos.

\vskip .2truein
\centerline {\it 4.2.~Break-up of Halos: Fall-in Method}

In the fall-in method we find all
of the bound $512^3$ DENMAX objects at an early epoch,
which we call the {\it tagging era}, and we find
the maximally bound particle from each of these halos.
We then grab the present day positions and velocities
of these particles and we add each one to our list of present day halos
only if it is a bound member of a massive halo with present day circular
velocity (defined within a radius 200 kpc) $\ge 350 \kms$.  Thus, we
break up the bound mass of large halos into several different objects
that were distinct entities at the tagging era.  Each such object is
assigned the circular velocity it had at the tagging era.  We also
retain the massive merged halo, unless the sum of masses of the added
halos exceeds its mass.

There are two arbitrary parameters in the method:
the circular velocity beyond which we break up the massive
halos and the epoch which we choose as the tagging era.
For the former we choose $V_{\rm circ}=350\kms$.  For larger
values we found excessive numbers of halos in Paper I.
We are admittedly forcing improved agreement with the observations.
For the tagging era we try $\sigma_8=0.2$ and $\sigma_8=0.3$.
There is no ideal, single epoch since galaxy formation
is a continual process.  WDEF eliminated this ambiguity
by finding every halo that ever formed and then by putting merging in
by hand.  We do not attempt to reproduce their procedure.

We show an example of the fall-in method at $\sigma_8=0.5$
in the lower left panel of Figure~7.  This example uses
$\sigma_8=0.2$ for the tagging era.  Ten objects
with $V_{\rm circ}\ge 250\kms$ fall into the massive halo
shown in the upper left panel.  When we use
$\sigma_8=0.3$ for the tagging era there are four halos
with $V_{\rm circ}\ge 250\kms$ that fall into this massive halo.
Larger values of $\sigma_8$ contain larger objects, but
many of these objects have already undergone merging.

We now examine the effect of the fall-in method on $\xi$ at $\sigma_8=0.5$.
We can see in Figure~7 that we introduce more pairs on
small scales (lower left panel) and that we give extra weight to the
massive halo (upper left panel).
We show $\xi$ after applying the fall-in method to CDM16 in the top panel
of Figure~9.  (The $M/{\cal L}$, method is discussed in the next section.)
Compare these results with the results without break-up in the top panel
of Figure~8.  We see better agreement of the slope with the observed slope
except on scales $<1$ Mpc.  We also see that we increase the
correlation length closer to the observed value.

The numbers of halos with no special treatment of the massive
halos at $\sigma_8=0.5$ are 1340 for $V_{\rm circ}\ge 200\kms$
and 737 for $V_{\rm circ}\ge 250\kms$.
The numbers from the fall-in method at $\sigma_8=0.5$
are 1934 (1706) for $V_{\rm circ}\ge 200\kms$ and 1022 (940)
for $V_{\rm circ}\ge 250\kms$ if we use $\sigma_8=0.2$ ($\sigma_8=0.3$)
as the tagging era.  There is a $\sim 30-40\%$ increase in the numbers
of halos using the fall-in method, showing that there has been a significant
amount of merging.

The fall-in method and the peak particles method produce nearly the same
shape of $\xi$ (see Figure~6).  However, the steepening of the slope occurs
at $r\sim 0.8$ Mpc using the fall-in method, compared with $r\sim 1.5$ Mpc
using the peak particles method.  Both produce the wrong shape.

Notice that the feature at $r\sim 1.5$ Mpc in
the bottom panel of Figure~4 is more prominent
than it is in Figure~9.  This is because the halos
are more extended using a $512^3$ DENMAX analysis in
a 100 Mpc box compared with using a $512^3$ DENMAX
analysis in a 51.2 Mpc box, as we discussed in Paper I.
The effectively coarser DENMAX allows more peripheral
particles.  These peripheral particles make the
halos bigger and introduce more pairs beyond $\sim 1.5$ Mpc.

The most important difference between the fall-in method
and the peak particles method is that the former requires far more
halos to get the same level of enhancement as the peak particle method.
This is to be expected because, as we demonstrated in $\S$ 3, the method
of peak particles oversamples the clusters and misses many field galaxies.
If galaxies cluster like dense dark matter halos, then it would seem
that the amount of clustering bias (i.e., the ratio of $\xi$ for
galaxies to that for the mass) must be less than predicted based
on peak particles.

\vskip .2truein
\centerline {\it 4.3.~Break-up of Halos: Mass-to-light Method}

As an alternative to the fall-in method for breaking up massive halos
we consider an ad hoc method designed to constrain the
mass-to-light ratio $M/{\cal L}$ of clusters.  We associate these galaxy
clusters with the massive merged halos and assign each such halo
the number of galaxies expected on average given a universal luminosity
function.  This method sacrifices all predictive power for cluster
$M/{\cal L}$'s, but we do not consider this a grave loss because we doubt
that any reasonable attempt can be made to estimate the luminosities of
galaxies in a cluster using a purely dissipationless simulation that
follows only the dark matter.  With the $M/{\cal L}$ method, we assume
only that the most massive halos should be associated with galaxy clusters
and that a specified $M/{\cal L}$ applies for all such clusters.  This
simple-minded prescription offers, at least, a useful foil for the fall-in
method.  Moreover, it allows us to vary the richness of clusters by varying
a single number, $M/{\cal L}$.  On the other hand, Ashman, Salucci,
\& Persic (1993) argue that observations of disk galaxies
imply a variable $M/{\cal L}$ which could
reduce the excess richness and numbers
of galaxies in clusters implied by hierarchical models assuming constant
$M/{\cal L}$.

Our procedure is straightforward.  We examine all halos with $V_{\rm circ}$
(defined at $R=200$ kpc comoving) exceeding $350\kms$.  For each such halo
we divide its total bound mass by a specified $M/{\cal L}$ (with ${\cal L}$
measured in the blue) to get the total blue luminosity in the cluster.
Ramella, Geller, \& Huchra (1989) find $M/{\cal L}\sim180h$ (in units of
${M}_{\odot}/{\cal L}_{\odot}$) for groups in the CfA2 survey.  Some
clusters are estimated to have values exceeding $500h$, but there is still
controversy among workers in the field.  Trimble (1987) gives a review.

We obtain a distribution of galaxies using the Schechter luminosity function
$\Phi({\cal L})$ with parameters $\Phi^{*}=1.56\times 10^{-2}h^3~{\rm
Mpc}^{-3}$, $M^{*}_{B_T}=-19.68-2.5\log_{10}h^{-2}$, and $\alpha=-1.07$
(Efstathiou, Ellis, \& Peterson 1988).
The total luminosity in a volume $V$ is
$${\cal L}_{\rm total}=V{\intop_0^\infty}{\cal L}\Phi({\cal L})d{\cal L}~.
\eqno(4.1)$$
The total number of galaxies in a volume $V$ with
a luminosity exceeding ${\cal L}$ is:
$$N(>{\cal L},V)=V{\intop_{\cal L}^\infty}\Phi({\cal L})d{\cal
L}~.\eqno(4.2)$$
Combining eq.~(4.1) and (4.2) and defining $x\equiv {\cal L}/{\cal L}_*$,
we get the total number of halos exceeding a luminosity ${\cal L}$
in a cluster with total light ${\cal L}_{\rm total}$:
$$N(>{\cal L},{\cal L}_{\rm total})={{\cal L}_{\rm total}\over {\cal L}_*}
{
{
\intop_{{\cal L}/{\cal L}_*}^\infty
{x^\alpha e^{-x}dx}
}
\over
{\Gamma(2+\alpha)} }~.\eqno(4.3)$$
Colless (1989) and Schechter (1976) inform us that
the same luminosity function works for rich clusters and for field
galaxies within the uncertainty of the data.

We now put the steps together.  We take the bound mass of one massive halo
(those with $V_{\rm circ}\ge 350\kms$) and we divide it by a specified
universal value of $M/{\cal L}$.  This gives us the total luminosity
emitted by the cluster: ${\cal L}_{\rm total}$.  We then add $N(>{\cal L},
{\cal L}_{\rm total})$ halos with luminosity exceeding ${\cal L}$ to the
big halo using eq.~(4.3).  We relate ${\cal L}$ to circular velocity
using the Tully-Fisher (Pierce \& Tully 1988) and Faber-Jackson relations
(using our fit from Faber et al. 1989; see paper I for details).
We assume 70\% spirals and 30\%
ellipticals; for the latter, the $V_{\rm circ}$ is corrected to $\sigma_1$
using eq.~(2.2) with $F=1$ (no significant difference occurs if we use
$F=1.1$).  The value of ${\cal L}$ corresponding to $V_{\rm circ}$ is
chosen so that the number in eq.~(4.3) exceeding
${\cal L}$ is the same as the number exceeding $V_{\rm circ}$.

When we add in halos using this mass-to-light method we
need to choose positions and velocities.  We do this
by randomly sampling the massive halos we are breaking up.
In other words, if the massive halo contains $N_h$ particles,
we generate a uniform random number $N_r$ from 1 to $N_h$ and
we use the present day position and velocity of particle
$N_r$.  We repeat this procedure for each added halo.
The break-up of the massive halo in the upper
left panel of Figure~7 is shown for the mass-to-light
method with random position sampling for
$M/{\cal L}=125$ in the lower right panel of Figure~7.
In Figure~9 we show $\xi$ at $\sigma_8=0.5$ for $M/{\cal L}=125$
where
we did this random sampling (long dashed curve),
and where we put all added halos on top of each
other at the locations of the original massive halos (dot-dashed curve).
The results agree at larger scales, but it is
essential to use the random sampling method to see
the effects from close pairs.  We use
random sampling for the remainder of this paper.

In Figure~9 we see that the mass-to-light method
produces results similar to the fall-in method and the peak
particles method.
We notice, however, that the slope on small scales
is steeper for the mass-to-light method than
for the fall-in method but comparable to
the peak particles method.  This is because
there are more galaxies added with $M/{\cal L}=125$
than with the fall-in method.  We quantify these numbers later.
By varying $M/{\cal L}$ we can test whether the spatial and velocity
statistics as well as numbers of halos are acceptable for a given model.
These quantities are not guaranteed to all work out satisfactorily even
with the freedom we allow ourselves in how massive halos are broken up.
We will see, on the contrary, that small-scale galaxy clustering and
velocities, combined with galaxy abundances and group multiplicities,
present serious difficulties for the CDM model.

\vskip .2truein
\centerline {\it 4.4.~Constraining $\sigma_8$ Using $\xi(r)$}

In this subsection we investigate the two-point correlation function
of simulated galaxies from our 100 Mpc high-resolution CDM simulation
CDM16($144^3$,100,85).  Galaxies are identified with DENMAX halos except
for the massive halos, which are split into several galaxies using
either the fall-in method or the mass-to-light method, with $M/{\cal L}$
a parameter that we can vary.  The purpose is to determine whether there
exists a normalization $\sigma_8$ such that $\xi$ for the simulated galaxies
matches the observations.  We vary the break-up procedure to determine the
sensitivity of our conclusions to this uncertain step.

Figure~10 shows results at three epochs $\sigma_8=0.5$, 0.7, and 1.0.
The shape of $\xi$ fails to match the observed $\xi$
in all cases and this must be considered to be a serious
shortcoming of the models.
At $\sigma_8=0.5$ the enhancement in $\xi$ is nearly
sufficient for $M/{\cal L}=50$ but the slope is
too steep on small scales.  Gott \& Turner (1979)
showed that the logarithmic slope $-1.8$
is valid down to at least scales of $\sim 10$ kpc with no indications
of any features on small scales.
The ${M/\cal L}=250$ case at $\sigma_8=0.5$ is not too
steep on small scales, but the correlation length
is only $\approx 7$ Mpc.  For ${M/\cal L}=500$ the
correlation length is $\approx 6$ Mpc and $\xi$ falls between
the no break-up case and the ${M/\cal L}=250$ case at small
scales at $\sigma_8=0.5$.  The no break-up
case at $\sigma_8=1$ is almost acceptable, but
the significant turnover on small scales does not
match the observed slope and the massive halos do not look anything like
observed clusters, i.e. they are single objects
rather than tens of objects.  At no epoch, with no treatment of
halo break-up, do the simulations match the observations.

We now examine the numbers of halos and the properties of
the clusters since these are central to further conclusions
regarding $\xi$.  In Table~2 we list the numbers
of halos with $V_{\rm circ}\ge 250\kms$.
The numbers are for the $(100~{\rm Mpc})^3$ volume.
The numbers without break-up are in the default column; we
also show numbers for $M/{\cal L}=50,$ 125, 250, and 500.

The observed number is less than 621 (563) for $V_{\rm circ}\ge 250\kms$,
assuming $\sigma_1=V_{\rm circ}/(\sqrt{3}/F)$ for ellipticals with $F=1$
($F=1.1$).  (Again, this is an overestimate because of the assumed
Faber-Jackson relationship; see Paper I.)  Even before breaking up the merged
halos the numbers are too large; breaking up the halos leads to an even
greater disagreement with observations.  The numbers for $M/{\cal L}=50$
and 125 are factors $\gsim 3-10$ too high!
Therefore, unless the mean $M/{\cal L}\gsim 1000h$ for typical groups,
we can safely rule out $\sigma_8\gsim 0.7$ just from the numbers shown
in Table~2.

The numbers for $M/{\cal L}=250$ at $\sigma_8=0.5$ are also high,
but may be consistent with observations within various uncertainties.
If we choose $M/{\cal L}\gsim 250$ we partially solve the high galaxy
abundance problem and the correlation function steepness problem,
but we do not raise the correlation length to the observed value.
Remember, WDEF found a factor of $\sim 3$ too many halos to yield
the required enhancement in $\xi$ at $\sigma_8=0.4$ (although they used
a 50 Mpc box), and we see in Figure~10 that the correlation length
falls short of the observed value at $\sigma_8=0.5$ for $M/{\cal L}=250$.
At later epochs we can solve the steepness problem using the catalogs
without breaking up the clusters, but then our simulations do not have
rich clusters like the real universe.

Finally, the numbers of galaxies from the fall-in method at $\sigma_8=0.5$
are comparable to the $M/{\cal L}=250$ numbers in Table~2 at $\sigma_8=0.5$.
This explains why $\xi$ at $\sigma_8=0.5$ looks markedly similar for the
fall-in method and for the mass-to-light method with $M/{\cal L}=250$.
This lends some support to our use of the mass-to-light method.  If
gaseous dissipation is able to preserve galaxies in clusters, even when
the dark matter halos merge (White \& Rees 1978; Katz \& White 1993;
Evrard et al. 1994), the CDM model might produce clusters of
galaxies with $M/{\cal L}\approx 250$, in not too violent disagreement
with the observations.

\vskip .3truein
\centerline {\bf 5.~CLUSTERS AND THEIR RICHNESS}

It is not enough for CDM or any other theory to predict the correct
two-point correlation function, pairwise velocity dispersions, and
galaxy abundances.  A successful theory must also predict the correct
richnesses, abundances, and mass-to-light ratios of galaxy groups and
clusters.  As we have noted, this test is difficult to make using a
purely gravitational N-body simulation without dissipation because of
the overmerging problem.  However, we showed in the preceding section
that a plausible scheme for undoing the overmerging is based on
assigning galaxies to massive merged halos in proportion to the mass.
Is it possible to do this with a reasonable value of $M/{\cal L}$ so
that the correct group multiplicity function (richness distribution)
is obtained?  Are there then too many clusters?

Let us recall first that the mean $M/{\cal L}$ for $\Omega=1$ is
$\sim 750$ for $h=0.5$.  On the other hand, most dynamical measurements
on cluster scales yield much values smaller by a factor of three or
more (see, e.g., Peebles 1986).  While velocity bias in clusters might
reduce the apparent $M/{\cal L}$ to acceptable values for $\Omega=1$,
there exist more direct mass measurements from X-ray emission for some
clusters (e.g., Hughes 1989).  We will therefore examine as well how
much mass is contained in our massive halos.

We compute the fraction of the total mass in our
$(100~{\rm Mpc})^3$ volume contained in massive halos
at $\sigma_8=0.5$, 0.7, and 1.0 using CDM16.
We accumulate the total bound mass
in all halos with $V_{\rm circ}$ (defined at 200 kpc comoving)
exceeding $350\kms$; these are the objects that we
have been breaking up in the previous section.
We find the percentage of mass contained
in these objects to be $19.2\%$ (267 objects) at $\sigma_8=0.5$;
$29.9\%$ at $\sigma_8=0.7$ (363 objects); and $39.9\%$ (420 objects)
at $\sigma_8=1.0$.  These numbers are excessive
when one recalls that only a few percent of galaxies
are in rich clusters; see Bahcall (1979) for a review.

Since these fractions are so high, we compute a few more
interesting numbers.
We compute the fraction of the mass contained
in objects at $\sigma_8=1.0$ using larger circular
velocity cut-offs.  For objects with
$V_{\rm circ}\ge 400\kms$ the mass fraction is $36.9\%$ (301 objects).
For objects with
$V_{\rm circ}\ge 500\kms$ the mass fraction is $31.5\%$ (170 objects).
Therefore, the amount of mass contained in very massive objects
is enormously high.  In Paper I we learned that
the cumulative mass fraction
converged with increasing mass resolution if
we imposed a distance cut.  Therefore, we
compute the mass fraction of objects above a given
circular velocity cut-off defined
at 500 kpc comoving,
and we accumulate only the bound mass within 500 kpc comoving.
At $\sigma_8=1$ for $V_{\rm circ}\ge 350\kms$
the mass fraction is $20.2\%$ (344 objects).
At $\sigma_8=1$ for $V_{\rm circ}\ge 500\kms$
the mass fraction is $15.0\%$ (156 objects).  The
numbers of objects are slightly less for cut-off radii
of 500 kpc comoving versus 200 kpc comoving because
many of the circular velocity profiles with
$V_{\rm circ}\sim 500\kms$ are actually falling slightly
at these scales.  However, for larger circular velocities
the profiles are still rising beyond 200 kpc comoving.

There is some uncertainty in defining the bound mass of
our massive halos.  However, even using a conservative
estimate we find that at least $15\%$ of the mass is contained
in very massive halos (with $V_{\rm circ}\ge 500\kms$)
at $\sigma_8=1$.  On the other hand, the percentage is not $100\%$ so we can
consider mass-to-light ratios $<750$ for our massive objects (remember that
$\Omega=1$ demands $M/{\cal L}=750$ on average) if $M/{\cal L}>750$ for
less massive objects.  Unfortunately for $\Omega=1$ CDM, this goes against
observations (Trimble 1987).  One cannot argue that the missing mass is far
outside galaxies in the CDM model (at least with $\sigma_8\gsim0.5$)
because more than half the mass is within 500~kpc from the center of a
halo (cf. Paper I).

Next, we consider the richness of our hand-made clusters
and we impose further constraints on the mass-to-light ratios;
the reader is reminded that the numbers of halos in our volume also
impose constraints (see Table~2).

Ramella, Geller, \& Huchra (1989; hereafter RGH) studied groups
of galaxies from the ${B(0)}\le 15.5$ CfA2 redshift survey.
Redshift space projection effects may bias some group properties
relative to groups selected using distance information as we do
in our simulations (Nolthenius \& White 1987).  Nevertheless, we
feel it is useful to present a preliminary analysis of the group
multiplicity function in the CDM model.
For our discussion in this section we convert all relevant quantities to
Zwicky magnitudes using ${B(0)}\approx {B_T}+0.29$ (Efstathiou, Ellis, \&
Peterson 1988).
We replicate our $(100~{\rm Mpc})^3$ volume using periodic boundary
conditions into a $(250~{\rm Mpc})^3$ volume.  We then
select a wedge corresponding to the CfA2 sky coverage:
right ascension range $8^h\le\alpha\le17^h$ and declination range
$26.5^\circ\le \delta < 38.5^\circ$.  We refer to this as the $12^\circ$ slice.
We assume $H_0=50\kms~{\rm Mpc}^{-1}$ and we impose
a distance cut of $R\le 240$ Mpc in our analysis.
We use actual galaxy positions rather than redshifts and
we impose an apparent magnitude limit of ${B(0)}\le 15.5$.  We assume
a Tully-Fisher relationship, see Paper I, converted to $M_{B(0)}$,
where we determine the circular velocities of the halos
from CDM16
at 200 kpc comoving.

We use DENMAX to identify all halos
with $V_{\rm circ}\ge V_{\rm circ}^{\rm MIN}=50\kms$; then
we use FOF to identify groups of halos
in our wedge after breaking up the massive halos ($V_{\rm circ}\ge
350\kms$) using the mass-to-light method.
We determine a FOF linking length, $l$ in Mpc, corresponding to
a given galaxy overdensity $\delta\rho/\rho$
given by $l^3=2/(n\delta\rho/\rho)$ (see, for example,
Frenk et al. 1988) where $n$ is the number density
of halos with circular velocity
exceeding $V_{\rm circ}^{\rm MIN}$ from our
original $(100~{\rm Mpc})^3$ volume.   We use FOF to identify
groups of halos after
breaking up the massive halos, but prior to imposing an apparent magnitude
limit.
Typical values of $l$, for $\delta\rho/\rho=80$,
range from 0.8 Mpc to 1 Mpc for
the various assumed values of $M/{\cal L}$ and $\sigma_8$.

We only identify groups with three or more members to
be consistent
with RGH.
RGH chose
a linking distance using a galaxy number density
estimated from
the observed Schechter luminosity function.
However, they
varied their linking length with redshift to account
for the sparse sampling of galaxies at large
redshift.
We avoid this difficulty by applying FOF with a fixed
linking length prior to applying an apparent magnitude
limit.  We then apply the apparent magnitude limit
to the resulting group catalog in a manner described below.

For field halos, i.e. those that are not in groups
with 3 or more members, we simply compute
$M_{B(0)}$ using the Tully-Fisher relationship, and
we remove those with ${B(0)}> 15.5$.
For the halos in groups we apply the following procedure.
If the group member is not created from the break-up of a massive
halo, then we eliminate it if ${B(0)}> 15.5$.
For group members that are created from the break-up of a massive
halo, we remove all of them and replace them by the
number of halos determined
from eq.~(4.3) for an assumed, universal $M/{\cal L}$.
The lower luminosity limit in eq.~(4.3)
is computed from $15.5-M_{B(0)}=5\log_{10}d+25.0$, where
$d$ is the distance to the group centroid in Mpc.  Note that
here we do not need to relate luminosity to $V_{\rm circ}$ in eq.~(4.3).
However, to be consistent with our use of $V^{\rm MIN}_{\rm circ}$,
we never allow ${\cal L}$ to fall below ${\cal L}_{\rm min}$ determined
from $V^{\rm MIN}_{\rm circ}$ using
the Tully-Fisher relationship.

The basic parameters in the group
finding algorithm are the galaxy overdensity $\delta\rho/\rho$ used
to determine the linking parameter,
the faint cut $V_{\rm circ}^{\rm MIN}$, the mass-to-light
ratio $M/{\cal L}$ used to break up the massive halos, and
the circular velocity cut-off above which we break up massive halos.
We discuss these four parameters here.

1) We report
results using $\delta\rho/\rho=80$, the middle
value considered by RGH, since we see
the same levels of variation with $\delta\rho/\rho$
as reported by RGH and our conclusions do not
depend critically on its value.

2) We
report results for $V_{\rm circ}^{\rm MIN}=50\kms$.
Our results do not depend sensitively on
$V_{\rm circ}^{\rm MIN}$ because the low mass galaxies
quickly fall out of sight.  For example,
in a case where we identify 1555 field galaxies in our
$12^\circ$ slice with an apparent magnitude limit,
only 233 have $V_{\rm circ}\le 125\kms$ and only 62 have
$V_{\rm circ}\le 75\kms$.  This is encouraging since
we found in Paper I that we had factors $\sim 2-3$
too many halos compared with the observations
for $V_{\rm circ}\lsim 125\kms$.  In a magnitude-limited survey
we would not be swamped by low mass halos.

3) We report results using $M/{\cal L}$=125, 250, and 500.
{}From a list of 36 groups, RGH found a median $M/{\cal L}$ of
$178 h=89$ for $h=0.5$.
We choose large values of $M/{\cal L}$ because, as we will
see, even $M/{\cal L}$=125 produces groups that
are too rich.

4) There is some arbitrariness to the value
of $V_{\rm circ}$ above which we break up the massive halos.
We report results using $V_{\rm circ}=350\kms$.  If we raise
this value we get too many isolated
massive halos (see Paper I) which are too big to
represent individual galaxies.
On the other hand, the numbers of halos added quickly approaches zero
below $V_{\rm circ}=350\kms$ for the $M/{\cal L}$ studied here.

The results from our simulations are shown in Table~3.
We report numbers from RGH for the full $12^\circ$ slice,
but we impose a redshift cut of $12000\kms$. RGH only
studied groups with centroids $\le 12000\kms$.
We report
numbers from the simulations for the full $12^\circ$ slice
for $R\le 240$ Mpc.
The table shows the number of groups, $N_{\rm groups}$, identified with
3 or more members, with 10 or more members, and with
20 or more members.  We also show the
number of galaxies, $N_{\rm galaxies}$, in the field, i.e.
those that are not in groups with 3 or more members.  We
estimate the number of CfA2 field galaxies within 12000 $\kms$ as follows.
The CfA2 catalog has 1766 galaxies and we estimate
from figure 1 in RGH that $\approx 100$ galaxies are beyond
12000 $\kms$.  RGH found 778 galaxies in groups with three or more members
and only a handful of these galaxies are beyond
12000 $\kms$.  Therefore, the number of field galaxies
within 12000 $\kms$ in the CfA2 catalog is approximately
$1766-778-100\sim 900$ galaxies.
The last entry in the table, $N_{\rm 1/2}$, is a richness statistic
defined later.  The reader is cautioned
that RGH estimate that $\gsim 30\%$ of the groups
with 3 or 4 members might be an artifact of projection effects.

We can draw several important conclusions from the results shown in
Table~3.  If we do not break up
the massive halos, then we do not have enough groups and
there are no groups with 10 or more members.
Therefore we need to break up our massive halos
if our simulated universe is to contain groups comparable
to the observed numbers!
In all cases we have too many field galaxies.
We demonstrated earlier that these are not dominated by
faint galaxies.  However, in Paper I we found that
we had the correct number of halos
with circular velocities between $150\kms$ and $350\kms$.
The reason for this discrepancy is that here we
apply only the Tully-Fisher relationship to the halos (i.e., we are
treating all halos as spirals)
rather than a combination of the Tully-Fisher
relationship and the Faber-Jackson relationship as we did
in Paper I.
Applying the Tully-Fisher
relationship to elliptical galaxies, which tend to
be the most massive halos, makes the halos appear
brighter than they really are.  On the other hand, most
of our group members result from the break-up of massive
halos where we do not need to assume a relationship
between circular velocity and luminosity.  Because of this problem,
we should give more emphasis to the richness of our groups
than to the apparent excess of field galaxies.  (We use the Tully-Fisher
relation only for field galaxies; massive halos are broken into galaxies
based on an assumed mass-to-light ratio.)

We can constrain
$M/{\cal L}\gsim 125=250h$ based on the number of
groups with 3 or more members and the total number of
galaxies in all groups with 3 or more members.
For $\sigma_8=1$, $M/{\cal L}$ must be $\sim 250=500h$.
In most cases, however, we still have too many rich groups with
50 or more members.  We should note that because the observed number
of groups with three or four members may be contaminated by projection
effects, the total numbers of objects in groups could be smaller by
$\gsim 30\%$ (see RGH).
This would lower the observed numbers in groups and, by definition,
raises the observed numbers in the field, although it does not solve
the problem of too many rich groups.

To further quantify the richness of our groups,
we compare the cumulative number of galaxies in groups
with the estimates from RGH for the CfA2 survey.
The cumulative number of galaxies in groups
is defined by RGH as:
$$N_{\rm galaxies}(\le N_{\rm members})\equiv {\sum_{N=3}^{
N=N_{\rm members}}}~N~N_g(N)~,\eqno(5.1)$$
where $N_{\rm galaxies}(\le N_{\rm members})$ is the total
number of galaxies contained in all groups with
three to $N_{\rm members}$ members and $N_g(N)$
is the number of groups containing $N$ members.
The results are shown in Figure~11 using $M/{\cal L}=125$, 250
and 500, and the no break up
cases, at $\sigma_8=0.5$, 0.7, and 1.0.
Figure~11 is computed for a $6^\circ$ slice (we divide
the numbers from our $12^\circ$ slice by two)
to compare with RGH using a $6^\circ$ (their figure~2).

We clearly see the dramatic shortcoming of the no break up cases
at all epochs.   When the massive halos are broken up we find
groups that are richer than observed by RGH;
the rise in the predicted cumulative galaxy number is also generally
slower than the results for the CfA2 survey
indicating that our group members are concentrated
in relatively larger groups.
A useful statistic is $N_{\rm 1/2}$ shown
in Table~3.  This is the value of $N_{\rm members}$
where the cumulative
number of galaxies in groups reaches 1/2 its maximum value.
The value of $N_{\rm 1/2}$
indicates that we need $M/{\cal L}\gsim 250$.  We can
rule out $M/{\cal L}=125$.
The remaining question is whether
or not nature can hide a lot of mass; this will be
an important consideration when we study velocities in the next section.

\vskip .3truein
\centerline {\bf 6.~HALO PAIR VELOCITY DISPERSIONS AND CLUSTERS}

We now consider constraints on $\sigma_8$ from
CDM16($144^3$,100,85) based on
pairwise velocity dispersions of the resolved halos.
We address the following questions:
What is $\sigma_{\rm p}$ for the halos without the break-up
of massive halos?
How do we assign velocities to halos added to the
massive halos, and what is the effect on $\sigma_{\rm p}$?
Is there a linear normalization of the $\Omega=1$
CDM model, $\sigma_8$,
when the pairwise velocity dispersions agree with the
observations?
\vfill
\eject
\vskip .2truein
\centerline {\it 6.1.~Constraining $\sigma_8$ Using $\sigma_{\rm p}(r)$
of Simulated Halos}

The pairwise velocity dispersions of the
halos from CDM16 without breaking up the massive halos
are shown in Figure~12
at $\sigma_8=0.5$, 0.7, and 1.0.
We define the circular velocities at 200 kpc comoving,
and we show results for $V_{\rm circ}\ge 100$, 150, 192 and 250
$\kms$.

The open symbols are the observed estimates
from the Davis \& Peebles (1983, hereafter DP) analysis of the
CfA ${B(0)}\le 14.5$ redshift survey.  The different
symbols are for different modeling parameters.
The best estimates are open circles with $1\sigma$ error
bars (shown as vertical lines).  The squares
are for a different set of modeling parameters, and
the triangles are results with three clusters removed.
The details are not important for our purposes; the
scatter is small compared with the $\sigma_8$ dependence
of $\sigma_{\rm p}$.  The results at $r\sim 10$ Mpc
are the least accurate because of
distortions from peculiar motions.
CDM16 has a Plummer softening of
65 kpc comoving which affects small scales.
For these reasons, we will focus our comparisons
at $r\sim 1$, 2.2, and 4.6 Mpc.

DP removed all galaxies with $M_{B(0)}>-18.5+5\log_{10}h=-20$.
If we convert this to the $B_T$ system and use
the Tully-Fisher relationship (see Paper I), this
corresponds to removing all halos with $V_{\rm circ} \lsim 175\kms$.
We study all halos with
$V_{\rm circ}\ge 150\kms$ and $V_{\rm circ}\ge 250\kms$. The former is
important
since the pairwise velocity dispersions increase with
increasing circular velocity cut-off and
simulated dispersions are higher than the observed estimates
at $\sigma_8\gsim 0.7$.
If we are to rule out any values of
$\sigma_8$, it is better
to be conservative.

Based on Figure~12, observational data constrains $\sigma_8
\lsim 0.7$.  The case $\sigma_8=0.5$ is an excellent
match to the observed data.  The results are
in reasonable agreement with the observed data at $\sigma_8=0.7$
for $V_{\rm circ}\ge 100$ and marginally for $V_{\rm circ}\ge 150\kms$.
The case $\sigma_8=1.0$ is ruled out; the pairwise velocity dispersions
are too high by factors $\sim 1.5$ for $r\gsim 1$ Mpc.
Note that this is true even though there is a velocity
bias of about a factor of two!

There are two important issues we need to consider.
We notice that the pairwise velocity dispersions are significantly
lower for the resolved halos than for the mass.
This velocity bias was discussed in $\S$ 2.4.
We also need to compare our results with Couchman \& Carlberg
(1992, hereafter CC) who investigated $\sigma_8\approx 1.0$
with a 2 million particle P$^3$M simulation in a 200 Mpc
box.  CC used a different definition to
normalize the linear CDM power spectrum; their
$b_\ell=0.8$ corresponds to $\sigma_8\approx 1.0$.
CC assumed $\Omega=1$, $H_0=50\kms~{\rm Mpc}^{-1}$,
and their particle mass is $2.65\times 10^{11}~{\rm M}_\odot$
compared with our particle mass of $2.3\times 10^{10}~{\rm M}_\odot$
for CDM16.

CC found a pairwise velocity dispersion for
halos with $M\gsim 2.1\times 10^{12}{\rm M}_\odot$ at 1 Mpc of $\sim 490\kms$
in agreement with our results for the lower circular velocity
cut-offs.  CC found a pairwise velocity dispersion for the
mass of $\sim 2300\kms/\sqrt{3}\approx 1325\kms$ at 1 Mpc; this is again in
agreement with our results.
CC did not report pairwise velocity dispersions on larger scales where
we find the disparity with the observations to be large.
CC also found that their halos have smaller two-point correlations
than the mass; this is in agreement with our results
presented in $\S$ 4.

We argued in $\S$ 4 that we need to break up massive halos
into clusters to remove the turnover of $\xi$ on small scales
and to enhance the correlation length, and in $\S$ 5
because clusters really exist in our universe.  We also
argued in $\S$ 2.4 that using the center-of-momentum
of resolved halos significantly reduced the pairwise
velocity dispersions compared with the mass since
a significant number of high
velocity particles are contained in a few massive halos.
For these reasons it is important to consider the effect
on $\sigma_{\rm p}$
of breaking up the massive halos
before further conclusions can be drawn.

We use the mass-to-light
method to break up the halos with $V_{\rm circ}\ge 350\kms$,
randomly sampling the positions and the velocities of
the massive halos to assign positions and velocities to the added halos.
The results are shown in Figure~13.
We see immediately that the pairwise velocity dispersion for the halos
now traces that for the mass.  We have introduced a significant
number of pairs with high velocity dispersions;
the added cluster members sample
massive halos which have high velocity dispersions.
These results indicate that the pairwise velocity
dispersions are too high at $\sigma_8=0.5$,
0.7, and 1.0 if we break up the massive halos.

CC did not report the high pairwise velocity dispersions
associated with clusters.  They found that merging
decreases the numbers of halos in high dispersion
regions, and they referenced Bertschinger \& Gelb (1991)
where we first discussed why this effect can significantly
reduce pairwise velocity dispersions.
However, CC did include a prescription
for preserving merged systems as distinct halos found by FOF, but they
commented that only $\sim 20\%$ of their
``galaxy precursors'' survive as distinct ``galaxies''.
A group analysis of the CC data, as we have done in $\S$ 5, is needed
to estimate their group multiplicity function.
Because our default catalogs (no break-up)
reveal $\sigma_{\rm p}$ in agreement with CC
at 1 Mpc, we suspect that they would see higher $\sigma_{\rm p}$
if they had the requisite group multiplicity function.

Before we can rule out any values of $\sigma_8$
we must examine lower circular velocity cut-offs.
We must also consider the possibility
that the velocity dispersions of galaxies in clusters
can be less than the velocity dispersions of the dark matter.
Finally, we must consider $M/{\cal L}=500$ at $\sigma_8=0.5$
which compares favorably with the observed properties of groups
of galaxies.
These tests are the focus of the next subsection.

\vskip .2truein
\centerline {\it 6.2.~Velocities of Added Cluster Members}

In the previous section we randomly sampled the velocities
of the particles in the massive halos to assign velocities to
the added cluster members.
An alternative method
is to use the one-dimensional velocity dispersion
of each massive halo as the rms for random numbers.

We compute $\sigma_1$ at
200 kpc comoving; $\sigma_1$ is
very flat at these scales (see Paper I).  We label
this quantity as $\sigma_1^{\rm (MH)}$;
MH is used to denote the original massive halo.
We label the $i^{\rm th}$ (for $i=x,~y,~z$) component of the
center-of-momentum velocity of the massive halo
as $v_i^{\rm (MH)}$.  We then compute
three gaussian random numbers, $r_i$, with
mean zero and a one-dimensional standard deviation $\sigma_1^{\rm (MH)}$
for each cluster member.
We define the velocity of the added cluster member as
$$v_i[{\rm cluster~member}]=v_i^{\rm (MH)}+\beta^{1/2} r_i~,\eqno(6.1)$$
for some constant $\beta\le 1$ discussed next.

The quantity $\beta$ is the ratio of ``galaxy temperature''
to the virial or gas temperature (see Sarazin 1988; Evrard 1990).
The ``galaxy temperature'' is a measure of the kinetic energy of the
galaxies and the gas temperature is directly related, in hydrostatic
equilibrium, to the gravitational potential well.
Observational estimates yield $\beta\sim 0.8$ (Evrard 1990) with
a range 0.4 to 1 (Table 2 of Sarazin 1988).

We show $\sigma_{\rm p}$ in Figure~14a and 14b using this method
to assign velocities to the added cluster members
with $\beta=1$, 0.8, and 0.25.  We use the mass-to-light
method to break up the halos with $V_{\rm circ}\ge 350\kms$.
We use $M/{\cal L}=250$ in Figure~14a,
$M/{\cal L}=500$ in Figure~14b, and in both cases we consider
halos with $V_{\rm circ}\ge 150\kms$;
these values, and $\beta=0.25$, are chosen specifically to
give low estimates of $\sigma_{\rm p}$.  We want
to know how much we need to ``push'' the parameters
to match the DP estimates of the pairwise velocity dispersions.
Admittedly, $\beta=0.25$ is far below the lowest
observational estimates ($\sim 0.4$ at best), and is only
shown as a final, futile attempt to save CDM!
The solid curves are for the mass.
The $\beta=1$ cases are comparable to the ${M/{\cal L}}=250$
cases using the random sampling method of Figure~13.
Here they are slightly lower because we show halos
with $V_{\rm circ}\ge 150\kms$ rather than for
$V_{\rm circ}\ge 250\kms$ used in Figure~13.
We conclude that even small $\beta$ cannot
save $\sigma_8\gsim 0.7$.  The case $\sigma_8=0.5$ still
has pairwise velocity dispersions that are high compared with
the observations for $M/{\cal L}=250$ and
the model requires
$\beta\lsim 0.25$ which is extremely small compared with observed
estimates.
The same conclusion holds for $M/{\cal L}=500$ at $\sigma_8=0.5$
except that the $\beta=0.25$ case is a reasonable match
to the observed pairwise velocity dispersions.  However, as mentioned
earlier,
$\beta=0.25$ is far below any observed estimate from
real clusters.

DP computed pairwise velocity dispersions, $\sigma_{\rm p}(r)$,
with and without the removal
of three clusters: Virgo, Coma, and A1367.  The triangles
in Figs.~14abc are the DP numbers computed without these three clusters, yet
their effect is small except for the 10 Mpc bin where the numbers
are unreliable.
However, the effect is not small for $\sigma_8=1$ CDM
as we now show.
The CDM model is plagued with far too many massive halos (see Paper I)
and too many rich clusters (see $\S~5$ of this paper).
In CDM16, we find 17 halos with $V_{\rm circ}>1000\kms$
at $\sigma_8=1$ (involving 395 galaxies with $V_{\rm circ}\ge 150\kms$
using the mass-to-light method with $M/{\cal L}=500$).
We find 2 halos with $V_{\rm circ}>1000\kms$
at $\sigma_8=0.7$ (involving 57 galaxies
with $V_{\rm circ}\ge 150\kms$, again using the mass-to-light method
with $M/{\cal L}=500$).
Last, we find no halos with $V_{\rm circ}> 1000\kms$ at $\sigma_8=0.5$.

We now compute $\sigma_{\rm p}(r)$ without the inclusion of
halos (or added members) with $V_{\rm circ}> 1000\kms$.
The results are shown in Fig.~14c for halos
with $V_{\rm circ}\ge 150\kms$, $\beta=0.8$, and $M/{\cal L}=500$ applied
to halos with $V_{\rm circ}\ge 350\kms$
(cf. Fig.~14b which includes all clusters).
The effect is substantial at $\sigma_8=1$; nevertheless,
$\sigma_{\rm p}(r)$ is still too large by a factor $\sim 2$ compared
with observations for $r\gsim 1$ Mpc.  In order to significantly reduce
$\sigma_{\rm p}(r)$ at $\sigma_8=1$ we must 1) assume a ridiculously
small $\beta$ and 2) remove an extreme number of rich
clusters.  Even so,
these effects are not strong enough to reconcile a $\sigma_8\gsim 0.7$
CDM universe with observed small-scale pairwise velocity
dispersions of galaxies.

As a final comment, we note some recent
work that is relevant to the formation of galaxies
in massive clusters in the CDM model.
Katz \& White (1993) (see also Evrard et al. 1994)
have performed a gas dynamical CDM simulation with
$\Omega=1$, $H_0=50 \kms~{\rm Mpc}^{-1}$,
and $\sigma_8=0.4$.  They simulated a volume of space
containing a massive halo found from a previous
dark matter only simulation. The object at $z=0$ ($z=1/a-1$;
$a=1$ at $\sigma_8=0.4$)
had a mass of $1.83\times 10^{14}~{\rm M}_\odot$ and
a circular velocity of $945\kms$; it formed
from the merging of two massive subclumps.
The gas dynamical simulation, assuming a 10 to 1
ratio of dark matter mass to gas mass, was evolved
to $z=0.13$.
During the course
of the simulation eight galaxies formed, but by
$z=0.13$ only four galaxies had survived the merging process.
(Each of these four galaxies had a cold gas mass exceeding
$1.9\times 10^{11}~{\rm M}_\odot$.)  Estimates
of $\beta$, with these limited statistics, were
$\beta\sim 1$ at $z=0.6$ and
$\beta\sim 0.4$ at $z=0.13$.

This work is interesting because it demonstrates
that some galaxies can survive the merging process.
However, this fact does not solve
the problems demonstrated in this paper.
Our mass-to-light method predicts that
a $1.83\times 10^{14}~{\rm M}_\odot$ object
should contain 4 halos with $V_{\rm circ}\ge 250\kms$
if $M/{\cal L}\sim 500=1000h$.  This is a factor
of 5 too few compared with
more typical $M/{\cal L}\lsim 200h$; see Trimble (1987).

Although the gas dynamical simulation of Katz \& White (1993)
demonstrated that some galaxies can survive the merging process
in a single massive halo, it did not demonstrate that
the $\Omega=1$ CDM model can successfully make clusters
of galaxies with reasonable mass-to-light ratios.
Furthermore, it did not demonstrate that CDM with
gas dynamics can solve the problems we have found
in this paper.  We leave open the possibility that
a full scale CDM simulation with gas dynamics might
significantly alter the distribution of luminous galaxies
compared with dark matter halos.  However, it is difficult
to imagine how this could avoid the problems associated
with having too many massive halos where galaxies are sure to form as
seen in the cosmological gas simulations of Katz et al. (1992).

On the other hand, we have found that $M/{\cal L}=500$ at $\sigma_8=0.5$
might solve some of the problems with the models.  The numbers of halos
and group properties were in good agreement with the observations.
However, the correlation length ($r_0\sim 6$ Mpc)
fell short of the observed value $r_0=10$ Mpc.
We found in this section that the velocities
for $M/{\cal L}\gsim 250$ at $\sigma_8=0.5$
are marginally
consistent with the observed pairwise velocity dispersions
and are in good agreement with the observed
pairwise velocity dispersions
for $\beta\lsim 0.25$ and $M/{\cal L}=500$.
If CDM is to survive on small scales,
nature must conspire to hide a lot of dark matter.
\vfill
\eject
\vskip .3truein
\centerline {\bf 7.~CONCLUSIONS}

There appears to be no
linear normalization of the power spectrum
for the $\Omega=1$ CDM model
that can simultaneously match the observed numbers, the
spatial clustering, and the pairwise velocity dispersion
of resolved dark matter halos.  The problems are especially
serious for the large amplitude ($\sigma_8\sim1.0$) implied by
the recent COBE-DMR anisotropy results.

We must break up the massive halos if our catalogs
are to contain groups of galaxies like the observed
universe.  If we study the models without breaking up
the massive halos, then we find that the two-point correlation
function turns over on small scales and the correlation
length is too small except for $\sigma_8\sim 1$ where
the turnover on small scales is particularly severe.
We also find that the pairwise velocity dispersions
constrain $\sigma_8\le 0.5$ despite the fact
that there is a velocity bias of a factor $\sim 2$.

We paid considerable attention to massive halos
which might represent groups of galaxies.
Breaking up these massive halos into groups of galaxies
removes the turnover of the two-point
correlation function on small scales and it increases
the correlation length on larger scales.
Unfortunately, the groups do more harm than good unless
we assume very high mass-to-light ratios.
They significantly increase the number of halos, they
give the wrong shape of the two-point correlation function,
they significantly increase the pairwise velocity dispersions, and
they make groups that are too rich for reasonable mass-to-light
ratios.  Our estimates constrain the models
to very high mass-to-light ratios $\sim 1000h$, although the
precise values are uncertain.  Factors such as 1) $\beta$,
2) how much of the bound mass should be used to estimate the
group luminosity (i.e. the mass out to a given distance from the group center),
and 3) variable mass-to-light ratios, all complicate the interpretation
of our estimated $M/{\cal L}$.  The combined uncertainty can be as much
as a factor of $\sim 2$.  However, there is increasing
evidence from X-ray studies of clusters
that dark matter is not hidden in the outskirts of
galaxy clusters (e.g. Sciama, Salucci, \& Persic 1992 and references
therein).

The problems associated with $\sigma_8\gsim 0.4$
are clear.  In agreement with White et al. (1987)
we found that we needed to restore halos in massive systems
to get the required two-point correlation length for
$\sigma_8=0.4$.  However, the fact that the model
then had a factor $\sim 3$ too many halos and produced
the wrong shape of the two-point correlation function
is a serious shortcoming of the model.  We also
studied models with $\sigma_8>0.5$.
In agreement with Couchman \& Carlberg (1992) we
found a velocity bias of a factor $\sim 2$ for $\sigma_8=1$.
However,
restoring the merged halos in massive systems which
have high velocity dispersions significantly increased the
pairwise velocity dispersions.  We can rule out $\sigma_8\gsim 0.7$;
even $\sigma_8=0.5$ required a ratio of galaxy to virial temperature
$\beta\lsim 0.25$ which is too small compared with observed estimates.
Removal of the most massive halos (with $V_{\rm circ}\ge 1000\kms$)
can reduce pairwise velocity dispersions, but the effect is too
little to save CDM with $\sigma_8=1$.

If we live in an $\Omega=1$ universe,
nature (or clever humans) must learn to hide large
amounts of dark matter.
Gas dynamical simulations
probably will not solve the problems we have found unless
our assumptions regarding sites of galaxy formation
and galaxy luminosities from the dark matter
are significantly wrong.

\vskip .3truein
\centerline{\bf ACKNOWLEDGEMENTS}
\vskip .1truein
This research was conducted using the Cornell National Supercomputer
Facility, a resource of the Center for Theory and Simulation in
Science and Engineering at Cornell University, which receives major
funding from the National Science Foundation and IBM
Corporation, with additional support from New York State
and members of its Corporate Research Institute.  We appreciate
the programming assistance of CNSF consultant Paul Schwarz.  We thank
Neal Katz for useful discussions on clusters, and David Weinberg and
Paul Schechter for stimulating discussions.
This work was supported by NSF grant AST90-01762 and in
part by the DOE and the NASA at Fermilab through grant NAGW-2381.

The DENMAX halo catalogs and full particle data sets are available
for the simulations used in this paper.  Interested workers may
send email to bertschinger@mit.edu.
\vfill
\eject
{\parindent 0pt
\centerline{\bf REFERENCES:}
{\pp Adams, F. C., Bond, J. R., Freese, K., Frieman, J. A.,
\& Olinto A. V. 1993, \hfil\break
\hbox{\hskip0.5truein} Phys. Rev. D, 47, 426}
{\pp Ashman, K. M., Salucci, P., \& Persic, M. 1993, MNRAS, 260, 610}
{\pp Bahcall, N. A. 1979, Ann. Rev. Astron. Astrophys., 15, 505}
{\pp Bardeen, J. M., Bond, J. R., Kaiser, N., \& Szalay, A. S. 1986,
ApJ, 300, 15}
{\pp Bertschinger, E. 1992, in New Insights into the Universe,
ed. V. J. Martinez, M. Portilla,\hfil\break
\hbox{\hskip0.5truein} \& D. Saez (Springer-Verlag: New York), 65}
{\pp Bertschinger, E. \& Gelb, J. M. 1991, Computers in Physics, 5, 164}
{\pp Bertschinger, E. \& Juszkiewicz, R. 1988, ApJ, 334, L59}
{\pp Carlberg, R. G. 1991, ApJ, 367, 385}
{\pp Carlberg, R. G. \& Couchman, H. M. P. 1989, ApJ, 340, 47}
{\pp Carlberg, R. G., Couchman, H. M. P., \& Thomas, P. 1990,
ApJ, 352, L29}
{\pp Cen, R. Y. \& Ostriker, J. P. 1992a, ApJ, 393, 22}
{\pp Cen, R. Y. \& Ostriker, J. P. 1992b, ApJ, 399, L113}
{\pp Colless, M. 1989, MNRAS, 237, 799}
{\pp Couchman, H. M. P. \& Carlberg, R.  1992, ApJ, 389, 453 (CC)}
{\pp Davis, M., Efstathiou, G., Frenk, C. S., \& White, S. D. M. 1985,
ApJ, 292, 371 (DEFW)}
{\pp Davis, M. \& Peebles, P. J. E. 1983, ApJ, 267, 465 (DP)}
{\pp Dubinski, J. \& Carlberg, R. G. 1991, ApJ, 378, 496}
{\pp Efstathiou, G., Bond, J. R., \& White, S. D. M. 1992, MNRAS, 258, 1p}
{\pp Efstathiou, G., Ellis, R. S., \& Peterson, B. A. 1988,
MNRAS, 232, 431}
{\pp Evrard, A. E. 1990, ApJ, 363, 349}
{\pp Evrard, A. E., Summers, F. J., \& Davis, M. 1994, ApJ, 422, 11}
{\pp Faber, S. M. \& Burstein, D. 1988, in Large-Scale Motions in
the Universe, ed.\hfil\break
\hbox{\hskip0.5truein} V. C. Rubin \& G. V. Coyne (Princeton: Princeton
University Press), p. 115}
{\pp Faber, S. M., Wegner, G., Burstein, D., Davies, R. L.,
Dressler, A., Lynden-Bell, D.,\hfil\break
\hbox{\hskip0.5truein} \& Terlevich, R. J. 1989,
ApJS, 69, 763}
{\pp Frenk, S. F, White, S. D. M., Davis, M., \& Efstathiou, G. 1988,
ApJ, 327, 507}
{\pp Gelb, J. M. 1992, M.I.T. Ph.D. thesis}
{\pp Gelb, J. M. \& Bertschinger, E. 1993, ApJ, submitted (Paper I)}
{\pp Gelb, J. M., Gradwohl, B., \& Frieman, J. 1993, ApJ, 403, L5}
{\pp Gott, J. R. \& Turner, E. L. 1979, ApJ, 232, L79}
{\pp Holtzman, J. A. 1989, ApJS, 71, 1}
{\pp Hughes, J. P. 1989, ApJ, 337, 21}
{\pp Kaiser, N. 1984, ApJ, 284, L9}
{\pp Katz, N., Hernquist, L., \& Weinberg, D. H. 1992, ApJ, 399, L109}
{\pp Katz, N., Quinn, T., Gelb, J. M. 1993, MNRAS, 265, 689}
{\pp Katz, N. \& White, S. D. M. 1993, ApJ, 412, 455}
{\pp Maddox, S. J., Efstathiou, G., Sutherland, W. J., \& Loveday, J. 1990,
MNRAS, 242, 43}
{\pp Nolthenius, R., \& White, S. D. M. 1987, MNRAS, 225, 505}
{\pp Park, C. 1990, MNRAS, 242, 59}
{\pp Park, C. 1991, MNRAS, 251, 167}
{\pp Peebles, P. J. E. 1986, Nature, 321, 27}
{\pp Pierce, M. J. \& Tully, B. 1988, ApJ, 330, 579}
{\pp Ramella, M., Geller, M. J., \& Huchra, J. P. 1989, ApJ, 344, 57 (RGH)}
{\pp Sarazin, C. L. 1988, X-ray Emission from Clusters of Galaxies
(Cambridge:\hfil\break
\hbox{\hskip0.5truein} Cambridge University Press)}
{\pp Saunders, W. et al., 1991, Nature, 349, 32}
{\pp Schechter, P. L. 1976, ApJ, 203, 297}
{\pp Sciama, D. W., Salucci, P., \& Persic, M. 1992, Nature,
358, 718}
{\pp Smoot, G. F. et al. 1992, ApJ, 396, L1}
{\pp Trimble, V. 1987, Ann. Rev. Astron. Astrophys., 25, 425}
{\pp White, S. D. M., Davis, M., Efstathiou, G., \& Frenk, C. S.
1987, Nature, 330, 451\hfil\break
\hbox{\hskip0.5truein}
(WDEF)}
{\pp Wright, E. L. et al. 1992, ApJ, 396, L13}
{\pp Vogeley, M. S., Park, C., Geller, M. J., \& Huchra, J. P. 1992,
ApJ, 391, L5}
}
\vfill
\eject
\parindent 0pt
\centerline{\bf FIGURE CAPTIONS:}

{\bf FIG.~1:} a) Linear theory predictions for
$\sigma_8$ as a function of box size $\lambda_{\max}$.
The power spectrum (Holtzman 1989, 5\% baryons)
is normalized so that $\sigma_8=1$
when $\lambda_{\rm max}\rightarrow \infty$.
b) Linear theory predictions for three-dimensional pairwise
velocity dispersions, $\sigma_v$,
as a function of box size, $\lambda_{\rm max}$,
and separation $r$, for four values of $\lambda_{\rm max}$.

{\bf FIG.~2:}
$\log_{10}\xi(r)$ (mass) versus $\log_{10} r$ where $r$ is measured
in comoving Mpc for various simulations in boxes ranging from 51.2 Mpc on a
side
to 400 Mpc on a side.
The five $128^3$ particle PM simulations ($R_{1/2}=280$ kpc, 51.2 Mpc box)
are averaged together (solid curves) with $1\sigma$ error bars.
The other simulations are
CDM11(128$^3$,102.4,560; dot-dashed curves),
CDM16(144$^3$,100,85; short-dashed curves), and
CDM10(128$^3$,400,2188; long-dashed curves).
Note that $\xi(r)$ is underestimated in the 51.2 Mpc boxes.

{\bf FIG.~3:}
$\sigma_{\rm p}(r)$
for the mass for the cases considered in
Figure~2.

{\bf FIG.~4:}
Average $\log_{10}\xi(r)$
for resolved halos with a) $V_{\rm circ}\ge 100\kms$,
b) $V_{\rm circ}\ge 200\kms$, and
c) $V_{\rm circ}\ge 250\kms$
from two $64^3$ particle, P$^3$M
simulations with $\epsilon=$40 kpc and 51.2 Mpc boxes.
The solid curves are for the mass
($1\sigma$ error bars are shown in the bottom panel)
and the solid lines are the observed $\xi$.
The results are shown for halos found with FOF ($l$=0.1; short-dashed curves)
and
FOF ($l$=0.2; long-dashed curves) and DENMAX (512$^3$ grid; dot-dashed curves).
In the bottom panel
we also show points (with $1\sigma$ error bars)
for catalogs with
the massive halos broken up.

{\bf FIG.~5:}
$\sigma_{\rm p}(r)$
at $\sigma_8=0.7$ from
CDM12($64^3$,51.2,52).
The solid curve is for the mass.  The dot-dashed curve is for the
mass
with the particles from the two largest halos removed.
The result for the DENMAX halos
with $V_{\rm circ}\ge 192 \kms$ is the long-dashed curve.
The result using the velocity of the maximally
bound particle in a halo rather than the center-of-momentum of the
halo is
the short-dashed curve.

{\bf FIG.~6:}
$\log_{10}\xi(r)$
for CDM16($144^3$,100,85)
where the ``galaxies'' are tagged from the initial density field
smoothed with a gaussian smoothing radius of $R_s=550$ kpc comoving
($\nu=2.6$ dot-long-dashed curves; $\nu=3.0$ short-dashed curves) and
880 kpc comoving ($\nu=2.0$ long-dashed curves; $\nu=2.5$ dot-short-dashed
curves).
The curved solid curve is for the mass
and the solid line is the observed $\xi$.

{\bf FIG.~7:}
A massive halo at $\sigma_8=0.5$ from CDM16.
a) The massive halo [with mass $2.1\times 10^{14}~{\rm M}_\odot$ and
$V_{\rm circ}(200~{\rm kpc})=666\kms$]
found by DENMAX; note that DENMAX fails to reveal
some visually distinct substructure.
b) Peak particles that are bound members of
the massive halo.  Circles are
for $V_{\rm circ}\ge 250\kms$ and squares are for
$200\kms\le V_{\rm circ}< 250\kms$.
c) Halos that formed at $\sigma_8=0.2$
that fell into the massive halo.
Circles are for $V_{\rm circ}\ge 250\kms$.
d) Halos added to the massive halo
assuming a mass-to-light
ratio of 125; positions are randomly sampled from
the massive halo; circles are for $V_{\rm circ}\ge 250\kms$
and squares are for $192\kms\le V_{\rm circ}< 250\kms$.

{\bf FIG.~8:}
$\log_{10}\xi(r)$ for DENMAX halos from CDM16.
The solid line is the observed $\xi$.
The curved line is for the mass.
Here there is no special treatment of massive halos.
The results are shown for $V_{\rm circ}\ge 100\kms$ (dot-long-dashed curves),
$V_{\rm circ}\ge 150\kms$ (short-dashed curves),
$V_{\rm circ}\ge 192\kms$ (long-dashed curves), and
$V_{\rm circ}\ge 250\kms$ (dot-short-dashed curves).

{\bf FIG.~9:}
$\log_{10}\xi(r)$ for DENMAX halos from CDM16 at $\sigma_8=0.5$.
We broke up massive halos with $V_{\rm circ}\ge 350
\kms$
using halos that fell in from $\sigma_8=0.2$ (dot-long-dashed curve) and
$\sigma_8=0.3$ (short-dashed curve).  We also show results assuming
a mass-to-light ratio of 125 where we randomly sample
the positions (long-dashed curve) and where we
put all added halos on top of each other (dot-short-dashed curve).
The solid line is the observed $\xi$.

{\bf FIG.~10:}
$\log_{10}\xi(r)$ for DENMAX halos with $V_{\rm circ}\ge 250\kms$ from
CDM16.  We break up the massive halos ($V_{\rm circ}\ge
350\kms$) using mass-to-light ratios: 50 (dot-long-dashed curves),
125 (short-dashed curves), and 250 (long-dashed curves).
The results without break-up are shown as dot-short-dashed curves.
The solid curves are for the mass and the solid
lines are the observed $\xi$.

{\bf FIG.~11:}
Cumulative number of galaxies in groups with
three or more members; see eq.~(5.1).
We break up the massive halos using the mass-to-light method with
$M/{\cal L}=125$ (top lines), 250, and 500; the higher curves
are for the smaller values of $M/{\cal L}$.  The lowest
curves are results
without breaking up the massive halos.
We assume $\delta\rho/\rho=80$ to determine the FOF
linking length.  The
results are for a $6^\circ$ wedge out to $R=240$ Mpc
with a magnitude limit ${B(0)}=15.5$.  (Note well, we divide
the numbers from a $12^\circ$ wedge in the simulation by two.
This explains why the jumps in the solid histograms are half
the value implied by $N_{\rm mem}$.)
The RGH results for a $6^\circ$ wedge
are shown for comparison as dashed histograms.

{\bf FIG.~12:}
$\sigma_{\rm p}$ for DENMAX halos from CDM16.
Here there is no special treatment of massive halos.
We show circular velocity cut-offs
of $100\kms$ (dot-long-dashed curves), $150\kms$ (short-dashed curves),
$192\kms$ (long-dashed lines), and 250 $\kms$ (dot-short-dashed curves).
The observed estimates
are shown as open symbols for various modeling parameters
from Davis \& Peebles (1983).
The solid curves are for the mass.

{\bf FIG.~13:}
$\sigma_{\rm p}$ for DENMAX halos with $V_{\rm circ}\ge 250\kms$ from
CDM16.  We break up the massive halos $(V_{\rm circ}\ge 350\kms)$
using the mass-to-light method:
$M/{\cal L}=50$ (dot-long-dashed curves),
$M/{\cal L}=125$ (short-dashed curves),
$M/{\cal L}=250$ (long-dashed curves).
(The positions and velocities
of the added halos are assigned using random sampling.)
The dot-short-dashed curves are without break-up.
The solid curves are for the mass.

{\bf FIG.~14:}
$\sigma_{\rm p}$ for DENMAX halos with $V_{\rm circ}\ge 150\kms$ from
CDM16.  The massive halos $(V_{\rm circ}\ge 350\kms)$
are broken up with a) $M/{\cal L}=250$
and b) $M/{\cal L}=500$.
The solid curves are for the mass.
The results without break-up are the dot-short-dashed curves.
The velocities of added members are generated from
the central velocity dispersions and $\beta$ (eq. (6.1)).
The $M/{\cal L}$ curves are for $\beta=1.0$ (dot-long-dashed curves),
$\beta=0.8$ (short-dashed curves), and
$\beta=0.25$ (long-dashed curves).
c)$\sigma_{\rm p}(r)$ for CDM16 halos with $V_{\rm circ}\ge 150\kms$,
removing those halos (and their progeny) with $V_{\rm circ}\ge 1000\kms$.
Halos with $350\kms\le V_{\rm circ}\le 1000\kms$ are broken up with
$\beta=0.8$ and $M/{\cal L}=500$.
\vfill
\eject
\end